\documentclass{IEEEtran}
\usepackage{cite}
\usepackage[switch]{lineno}
\usepackage{amsmath,amssymb,amsfonts}
\usepackage{subfigure,dcolumn}
\usepackage{siunitx}
\usepackage{csquotes}
\usepackage{mhchem} 
\usepackage{url}      
\usepackage{hyperref}  
\usepackage{graphicx}
\usepackage{textcomp,nicefrac}
\def\BibTeX{{\rm B\kern-.05em{\sc i\kern-.025em b}\kern-.08em
T\kern-.1667em\lower.7ex\hbox{E}\kern-.125emX}}
\begin{document}
\title{A MIDAS-based Data Acquisition System for Gaseous Detectors} 

\author{Yuanchun Liu, Tao Li, Yu Chen, Ke Han, Leyan Li, Shaobo Wang, and Wei Wang
\thanks{Manuscript received xx xx xx; accepted xx xx xx. Date of publication xx xx xx; date of current version xx xx xx.}
\thanks{This project is supported in part by grants from the National Key R\&D Program of China~(No. 2024YFF0727503), the Innovation Funds of CNNC (Lingchuang Fund No. CNNC-LCKY-2025-051), the Natural Science Foundation of
Shanghai~(Nos. 24ZR1437100 and 25ZR1402223), and the Natural Science Foundation of Sichuan~(No. 2026NSFSC0757).  (Corresponding authors: Tao Li and Shaobo Wang.)}
\thanks{Yuanchun Liu is with the INPAC and School of Physics and Astronomy, Shanghai Jiao Tong University, MOE Key Lab for Particle Physics, Astrophysics and Cosmology, Shanghai Key Laboratory for Particle Physics and Cosmology, Shanghai 200240, China.}
\thanks{Tao Li is with the SPEIT~(SJTU-Paris Elite Institute of Technology), Shanghai Jiao Tong University, Shanghai 200240, China; and Shanghai Jiao Tong University Sichuan Research Institute, Chengdu 610213, China (e-mail: taoli@sjtu.edu.cn).}
\thanks{Yu Chen is with School of Physics, Sun Yat-Sen University, 135 Xingang Xi Road, Guangzhou, 510275, China.}
\thanks{Ke Han is with INPAC and School of Physics and Astronomy, Shanghai Jiao Tong University, MOE Key Lab for Particle Physics, Astrophysics and Cosmology, Shanghai Key Laboratory for Particle Physics and Cosmology, Shanghai 200240, China; Shanghai Jiao Tong University Sichuan Research Institute, Chengdu 610213, China; and Jinping Deep Underground Frontier Science and Dark Matter Key Laboratory of Sichuan Province, Xichang 615000, China.}
\thanks{Leyan Li is with the SPEIT~(SJTU-Paris Elite Institute of Technology), Shanghai Jiao Tong University, Shanghai 200240, China.}
\thanks{Shaobo Wang is with the INPAC and School of Physics and Astronomy, Shanghai Jiao Tong University, MOE Key Lab for Particle Physics, Astrophysics and Cosmology, Shanghai Key Laboratory for Particle Physics and Cosmology, Shanghai 200240, China; SPEIT~(SJTU-Paris Elite Institute of Technology), Shanghai Jiao Tong University, Shanghai 200240, China; Shanghai Jiao Tong University Sichuan Research Institute, Chengdu 610213, China; and Jinping Deep Underground Frontier Science and Dark Matter Key Laboratory of Sichuan Province, Xichang 615000, China (e-mail: shaobo.wang@sjtu.edu.cn).}
\thanks{Wei Wang is with School of Physics, Sun Yat-Sen University, 135 Xingang Xi Road, Guangzhou, 510275, China; and Sino-French Institute of Nuclear Engineering and Technology, Sun Yat-sen University, Zhuhai 519082, China}}

\maketitle

\begin{abstract}
We present a data acquisition~(DAQ) software based on the MIDAS framework, specifically for gaseous detectors to support the detector deployments and applications.
It implements a comprehensive suite of functions, including parameter configuration, data acquisition, decoding, and storage, alongside web-based operation and real-time monitoring capabilities.
We establish a fully unified workflow spanning data acquisition to offline analysis, enabling real-time visualization of signal waveforms and energy spectra. 
The system has been successfully deployed in the PandaX-III experiment, which utilized a high-pressure gaseous detector to search for neutrinoless double beta decay. 
Its performance and stability have been validated through tests involving two distinct electronics setups and joint commissioning with the detector.
\end{abstract}


\section{Introduction}
\label{sec:introduction}
\IEEEPARstart{G}{aseous} detectors, renowned for their superior particle imaging capabilities, are widely employed in particle, nuclear, and astroparticle physics experiments.  Specifically, gaseous Time Projection Chambers~(TPCs)~\cite{b1} can precisely record the three-dimensional spatial coordinates and energy depositions at each sampling point along a particle trajectory, providing rich event-related information which offers unique advantages in track topology reconstruction, a key capability for signal and background discrimination in experiments~\cite{b2, b3}.
In rare event searches, gaseous TPCs are extensively utilized for Migdal effect measurements~\cite{b4}, dark matter searches~(e.g., MIMAC~\cite{b5}, CYGNO~\cite{b6}), and neutrinoless double beta decay~($0\nu\beta\beta$) searches~(e.g., NEXT~\cite{b7}, N$\mathrm{\nu}$DEx~\cite{b8}, PandaX-III~\cite{b9}). 
However, the dual requirements of imaging accuracy and detector size pose significant challenges to the electronics and data acquisition~(DAQ) systems.

To address these challenges, we have developed a DAQ software based on the MIDAS~(Maximum Integrated Data Acquisition System) framework~\cite{b10}. 
Initially introduced for the Pion Beta experiment~\cite{b11} at the Paul Scherrer Institute~(PSI), MIDAS has since been widely adopted by various experiments, including T2K~\cite{b12}, DEAP-3600~\cite{b13}, and GRIFFIN~\cite{b14}, due to its robustness and versatility. 
The DAQ software follows the logic of the firmware and embedded software of the back-end module, enabling efficient communication across the entire electronics hierarchy. 
Its various functions include trigger mode settings, counter measurements, timestamp recording, and the organization of digitized samples from all channels into a structured event format for data collection.

The DAQ software fully leverages the built-in functions of the MIDAS framework, including the User Interface~(UI), online analysis, slow control, and data storage. 
While it possesses a certain degree of universality for gaseous detectors, it is also customized to meet the specific requirements of gaseous detector-based experiments. 
For instance, it adopts a web-based interface to enable real-time monitoring of experimental status, automatic log recording, and timely alarm notifications. 
Simultaneously, it employs an online database architecture, which allows convenient configuration of electronic parameters by modifying variable key values via the web interface. Furthermore, the software interfaces with the REST data analysis framework~\cite{b15}, realizing seamless integration of data acquisition, event reconstruction, and analysis so that the digitized waveforms can be visualized in real time during the acquisition process.

As a typical application, we conducted a series of joint tests with the PandaX-III experiment~\cite{b16, b17}, which aims to search for the $0\nu\beta\beta$ decay of the $^{136}$Xe isotope at the China Jinping Underground Laboratory~\cite{b18}. 
The readout plane of the PandaX-III detector consists of a tessellation of 52 Micromegas~(Micro-Mesh Gaseous Structure) modules~\cite{b19, b20}, each with a size of 20$\times$20 cm$^2$, totaling 6656 readout channels. 
This large-scale, multi-channel readout system not only imposes stringent requirements on the DAQ system, including real-time synchronized readout across multiple channels, flexible configuration of electronic parameters, reliable monitoring of experimental operating conditions, and seamless integration with subsequent data processing workflows, but also serves as an ideal testbed for validating the performance of our DAQ software.

This paper is organized as follows: The performance test platform of the MIDAS-based DAQ software, the PandaX-III experiment, is described in Section~\ref{sec.II}. 
The architectural design and data format of the DAQ software are presented in Section~\ref{sec.III}. 
The functional overview of the software is displayed in Section~\ref{sec.IV}, and performance tests are detailed in Section~\ref{sec.V}. Finally, we provide the conclusions in Section~\ref{sec.VI}.

\section{PandaX-III experiment}\label{sec.II}

The PandaX-III experiment utilizes a high-pressure gaseous TPC to search for the $0\nu\beta\beta$ signals.
The detector vessel of the PandaX-III experiment is a stainless steel container with a volume of 6.3~m$^3$ that contains about 140~kg of gaseous xenon at 10~bar.
The field cage is constructed from an acrylic structure and flexible printed circuit boards, enclosing an internal sensitive volume with a 1.6~m diameter and 1.2~m height.
The charge readout plane of the PandaX-III TPC is composed of 52 Thermal Bounding Micromegas~(TBMM) modules~\cite{b21, b22} for amplifying and readout electrical signals.
Each module comprises 128 readout strips with a pitch of 3~mm.
The stainless steel mesh is supported by pillars along the readout strips, forming a 100~\si{\micro\meter} amplification gap.
The signal from the mesh, which is the summation of all strip signals, can also serve as an external trigger. 
Consequently, 6656 strip channels are required for the electronics system.

\begin{figure}
    \centering
    \includegraphics[width=1.\linewidth,trim=125 50 125 45,clip]{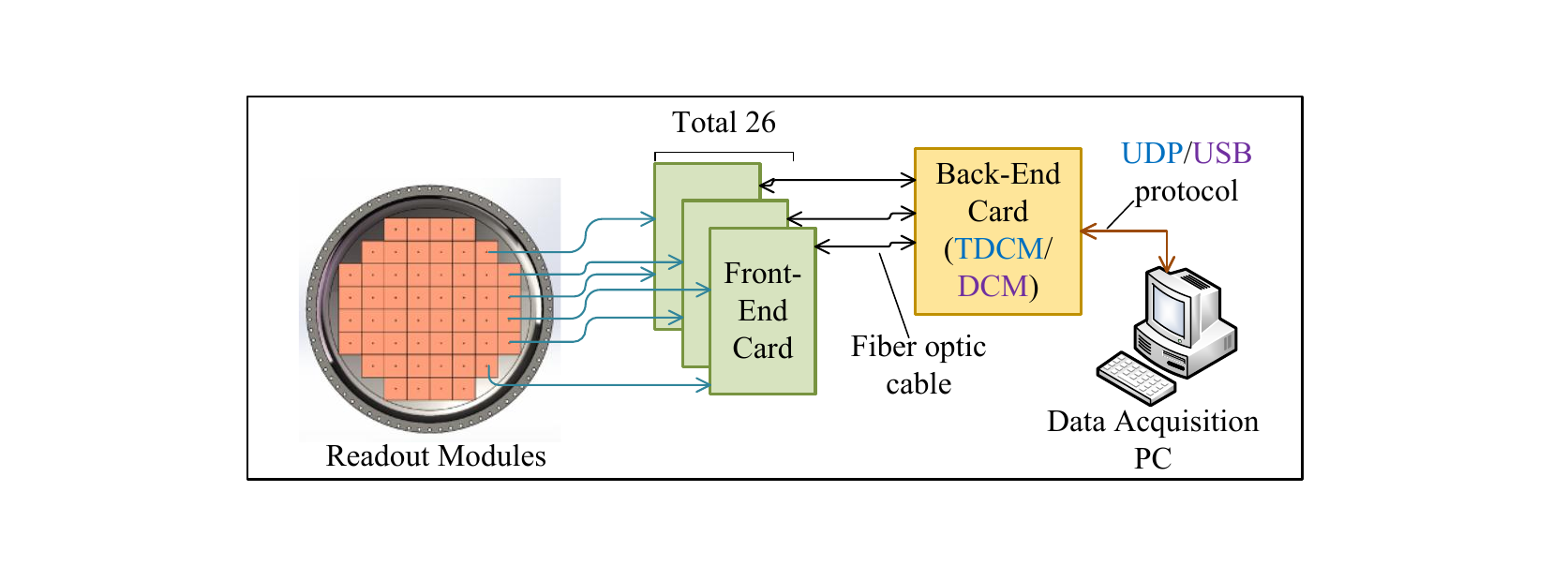}
    \caption{The structure of electronics system of PandaX-III, consisting of FEC, TDCM/DCM, and DAQ.}
    \label{fig:p3_daq_system}
\end{figure}

\begin{figure*}[htbp]
    \centering  
    \subfigure[FEC]{
        \includegraphics[width=0.3\hsize]{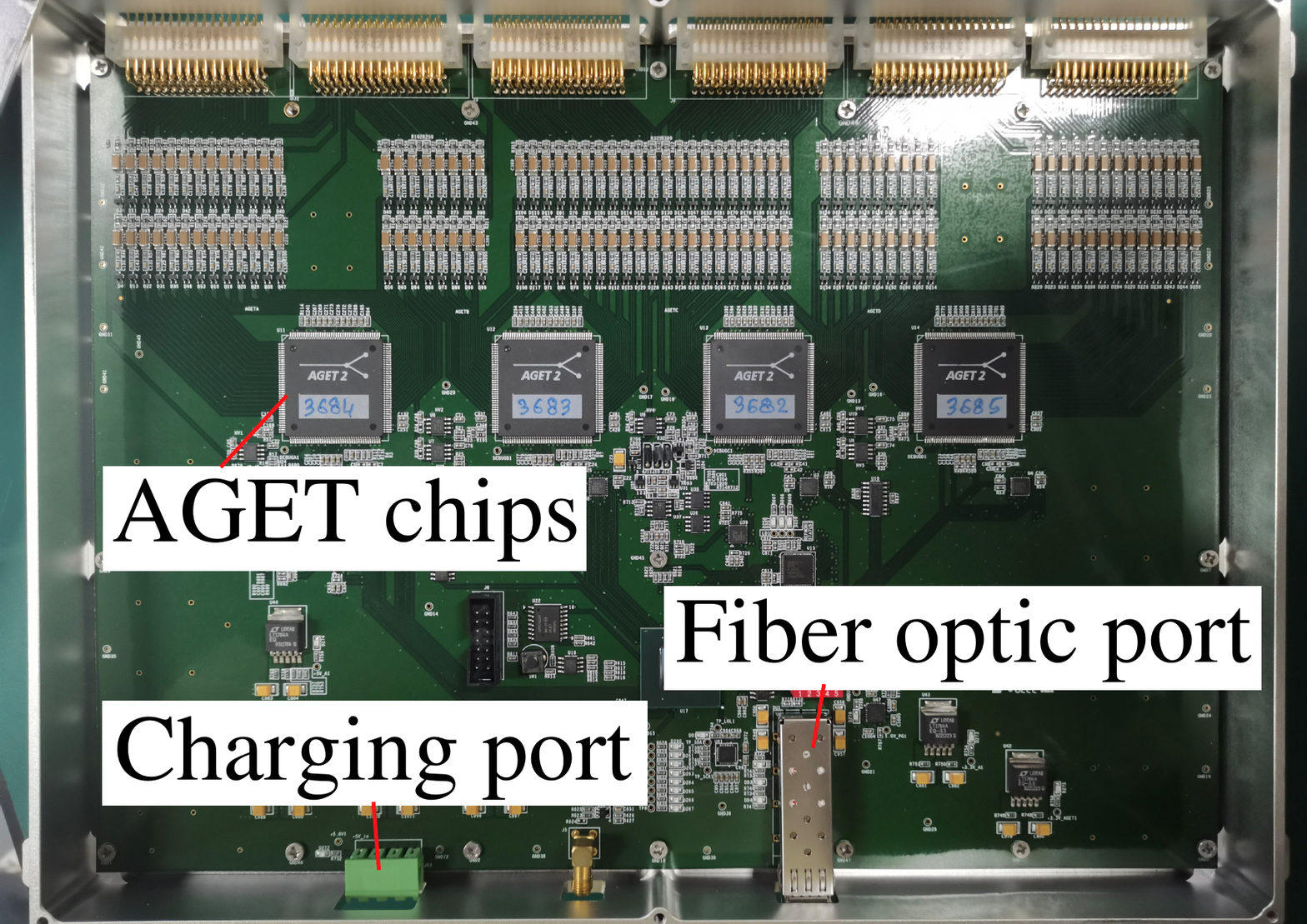}
        \label{fig::FEC} 
    }
    \subfigure[TDCM]{
        \includegraphics[width=0.3\hsize]{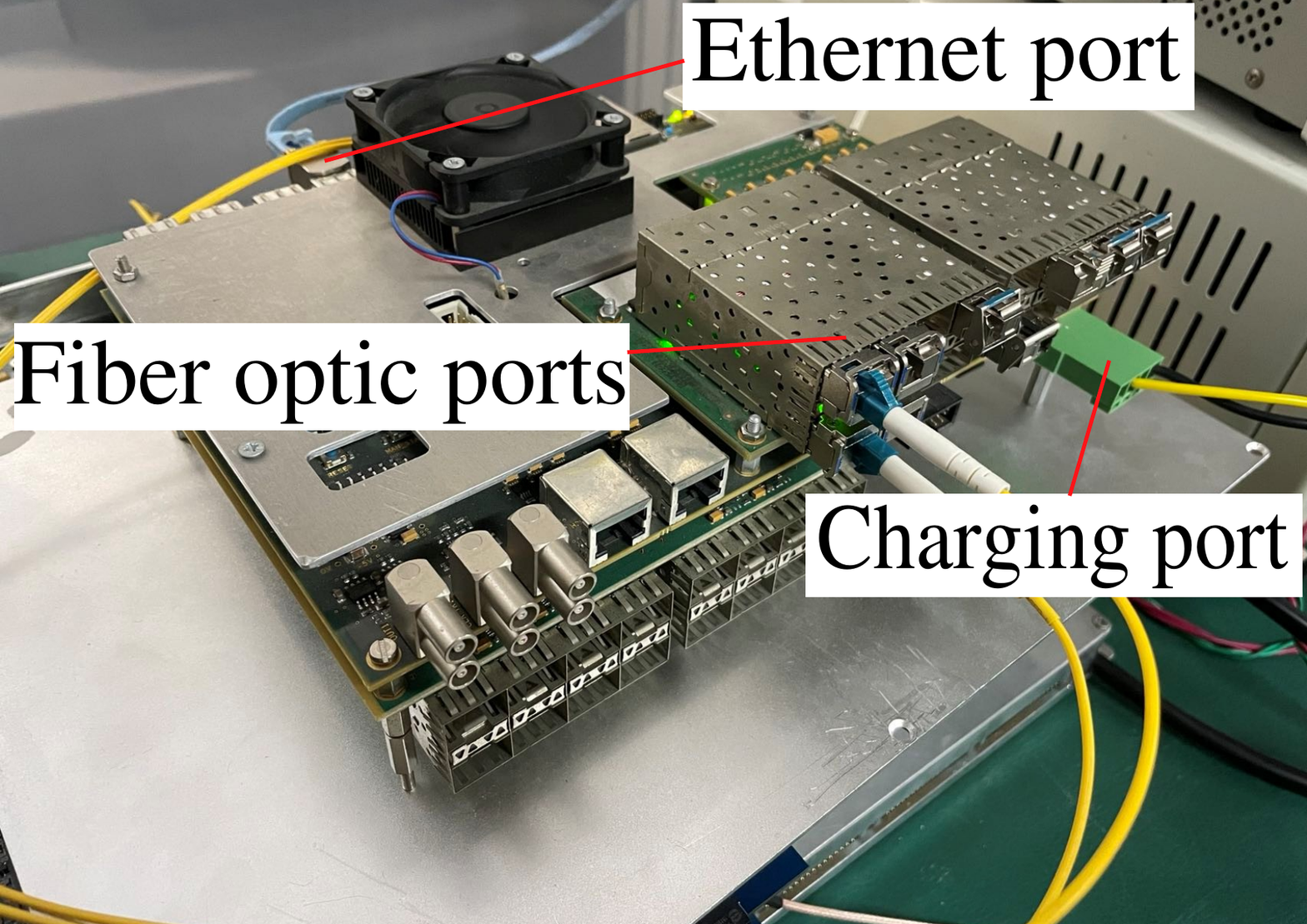}
        \label{fig::TDCM} 
    }
    \subfigure[DCM]{
        \includegraphics[width=0.3\hsize]{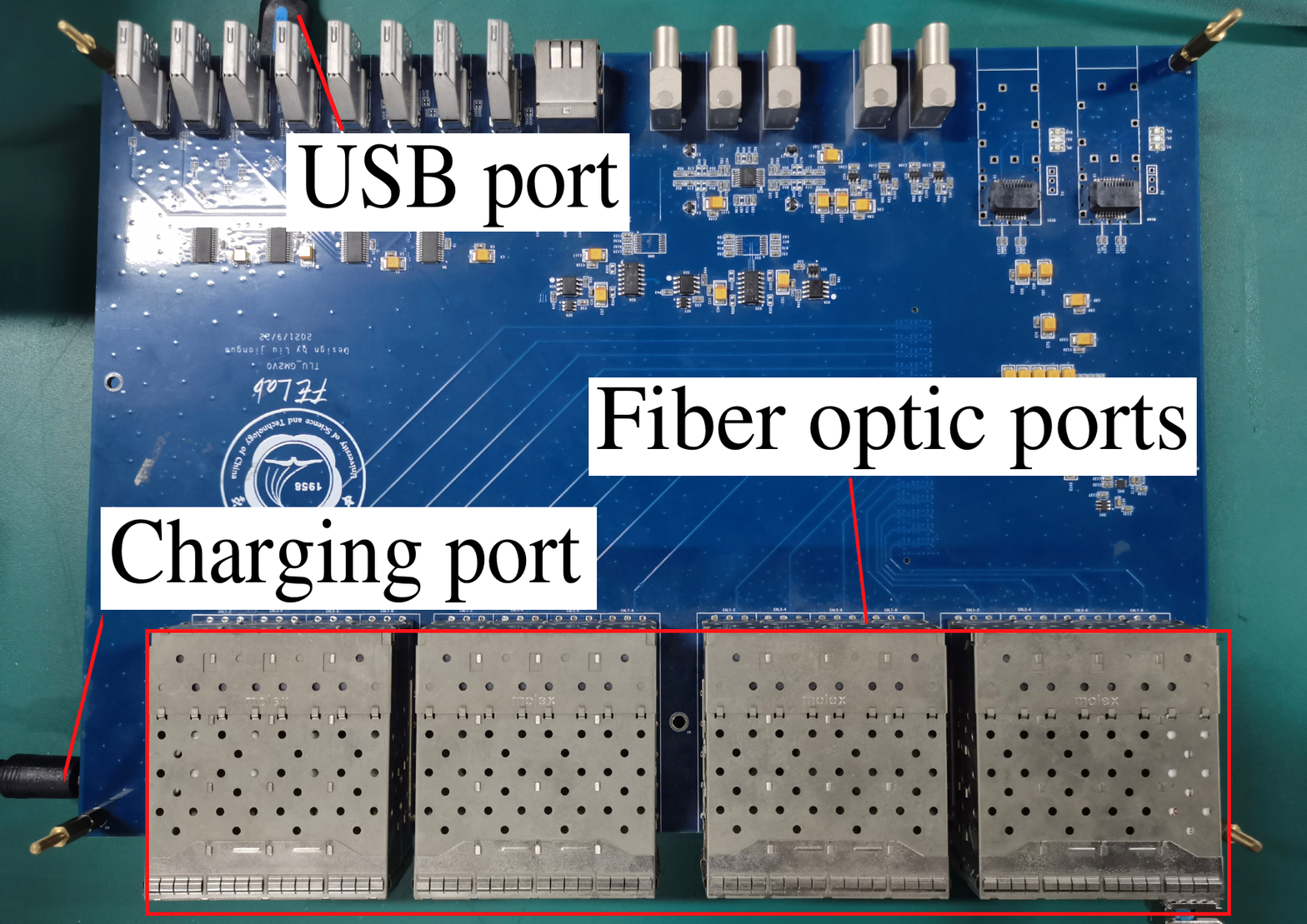}
        \label{fig::DCM}
    }
\caption{The two different electronics systems of PandaX-III: The front-end electronics FEC~(a), and the back-end electronics TDCM~(b) and DCM~(c) of the electronic system in the PandaX-III experiment.}
\label{electronics system}
\end{figure*}

The electronics system includes 26 Front-End Cards~(FEC)~\cite{b23} and 1 Back-End Card~(BEC), as illustrated in Fig.~\ref{fig:p3_daq_system}. 
At the lowest level, the signal readout is performed by 26 FECs for all Micromegas modules. 
Each FEC is equipped with 4 AGET~\cite{b24} chips, each of which has 68 readout channels~(64 data channels and 4 noise channels), as shown in Fig.~\ref{fig::FEC}. 
One FEC can handle the signal readout from two TBMM modules. 
The FEC offers four selectable dynamic ranges: 120~fC, 240~fC, 1~pC, and 10~pC.
The waveform sampling rate can be adjusted between 1~MHz and 100~MHz. 
Additionally, the FECs read out the mesh signals and generate individual "Mesh-trigger" signals.
The FECs receive signal pulses from the TBMM modules, integrate the charges, and digitize them under the control of the trigger. 
The digitized data and status information are packed and transmitted to the BEC using a user-defined serial protocol with an optical link. We have two types of BEC, TDCM~(Trigger and Data Concentration Module)~\cite{b25} is developed by CEA Saclay, France, and DCM~(Data Concentration Module)~\cite{b26} is developed by USTC, China. 
The hardware interface between the FEC and the BEC utilizes Small Form-factor Pluggable transceivers that comply with the Multi-Source Agreement.

At the intermediate level, the BEC~(TDCM and DCM), as shown in Figure~\ref{fig::TDCM} and Figure~\ref{fig::DCM}, is responsible for collecting event data packets and status information from the FECs and sending out commands and configuration data to the FECs.  
The BEC handles trigger control and data transmission for all 26 FECs, with the capability to accommodate up to 32 FECs.
Communication between the BEC and FEC boards is achieved through fibre optic cables for point-to-point transmission. 
The interaction between the BEC and the DAQ software is established through two methods: Gigabit Ethernet cables using the UDP protocol and Universal Serial Bus~(USB) communication.
The BEC performs various functions, including front-end and back-end clock synchronization, configuration and command feedback for the electronics, front-end data readout, as well as slow control and monitoring of the electronics.
A credit score mechanism is used to address potential data loss during transmission.
A precision crystal oscillator with a frequency of 25 or 50~MHz is utilized in the BEC. 
The clock signal and other control signals, such as the event trigger, timestamp reset, and event counter reset, are distributed from the BEC to the corresponding FECs.
The TDCM is equipped with the $multiplicity$ feature to cater to different data acquisition requirements in experiments.
For instance, certain multi-channel triggered physical events can be filtered by configuring the $multiplicity$ parameter, which specifies the number of triggering channels.

At the top of the hierarchy, a PC farm running the DAQ software receives event data from the BEC via USB~3.0 protocol or Gigabit Ethernet~(GbE) links using the UDP protocol.
To handle the large number of channels and the electronic system, the DAQ is developed based on the MIDAS framework, offering a user-friendly web application for configuring the electronic functionalities of the FEC and BEC, as well as controlling the data acquisition system.
The received data frames will be stored on disk in a specific format for subsequent data analysis.
More details about the DAQ software will be provided in the following section.

\section{DAQ Architecture}\label{sec.III}

The hardware described above requires DAQ software capable of dynamically managing protocol translation, trigger synchronization, and real-time data monitoring.
To address these requirements, a DAQ software architecture based on the MIDAS framework was developed. 
MIDAS is a universal event-based data acquisition framework designed for physics experiments, which provides three default applications: \enquote{odbedit}, \enquote{mhttpd}, and \enquote{mlogger} for parameter configuration, web interaction, and data storage, respectively.
In addition to that, the DAQ software provides two custom applications: \enquote{CmdProc} and 
\enquote{DataFlow}, which are primarily responsible for electronics configuration and data collection for the PandaX-III experiment, respectively.
These applications mentioned above in the DAQ software are summarized as follows: 
\begin{itemize}
    \item \enquote{odbedit}: MIDAS builds an ODB with a tree-based structure to manage hardware and experimental configurations. The core ODB trees~(e.g., System, Runinfo, Experiment) pre-defined during initialization store critical information, such as system state, runtime parameters, and equipment settings. ODB is editable and can be finely customised for DAQ systems through parameter modification or directory extensions.
    \item \enquote{mhttpd}: Through the mhttpd application, users can connect to the DAQ system using any browser and control the experiment via a set of interactive buttons for parameter configuration and experiment control, in addition to providing hyperlinks to dedicated subpages.
    \item \enquote{mlogger}: The purpose of mlogger is to save data from the experiment onto disk. Before commencing data acquisition, mlogger must be run within the directory designated for storing experimental data on the server.
    \item \enquote{CmdProc}: CmdProc is an application for monitoring command configuration and receiving feedback from the backend electronics. Once launched, the application provides a terminal interface for command input to query the electronics status and converts the hexadecimal output of the devices into human-readable text.
    \item \enquote{DataFlow}: Dataflow focuses only on correctly receiving and processing data frames sent by the back-end electronics to the DAQ software. Upon execution, the terminal displays real-time metrics including event count, sampling rate, and storage velocity.   
\end{itemize}

To dissect the operational mechanism of the above modular architecture, this section proceeds with five key aspects: Overall workflow, Communication with the electronics, Web interface, Data format, and Analytical Framework.

\subsection{Overall workflow}

\begin{figure*}[htbp]
    \centering
    \includegraphics[width=\hsize,trim=20 150 20 100,clip]{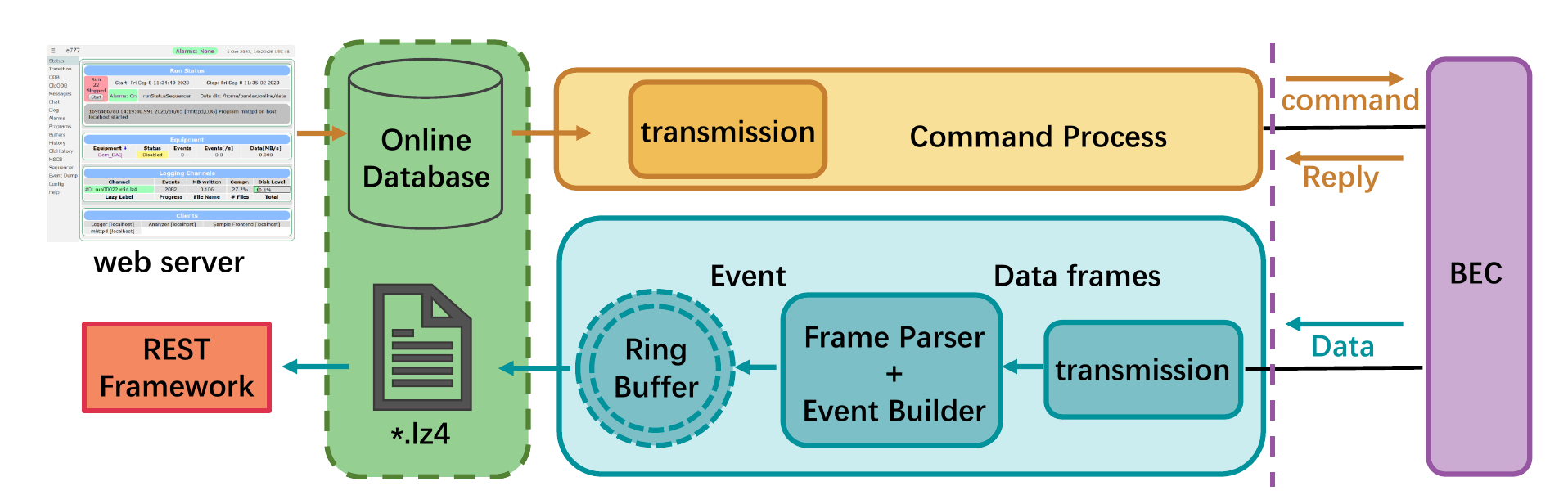}
    \caption{Schematic diagram of the main workflow for the DAQ software, including command transmission from the web client to the backend, data exchange between the backend and the DAQ device, as well as frame parsing and event reconstruction processes.}
    \label{DAQ_structure}
\end{figure*}

The DAQ has two data flows that are responsible for the message transmission and data collection, with the architecture schematically illustrated in Fig.~\ref{DAQ_structure}.
The received data frames are classified based on their prefixes, which determine the corresponding workflow.

In the message transmission, the DAQ enables user-defined configuration by modifying the ODB via a web interface.
The configuration parameters are forwarded to the DAQ system, which applies them to the electronics.
The resulting feedback from the electronics is collected and displayed on the web interface.
Moreover, the ODB supports simultaneous communication with multiple hardware devices, providing real-time status monitoring and independent parameter configuration for each device.
Information returned from the electronics is displayed in the terminal through the \enquote{CmdProc} application.

Data collection focuses on processing and storing the raw data streams from the electronics.
Specifically, the DAQ receives the real-time raw data sent by the back-end electronics.
Data frames are parsed into header, trailer, and waveform data segments.
The header contains information such as the event size, timestamp, and index number.
New events are reconstructed based on header information and the entire waveform data.
Following reconstruction, events are stored on the server in LZ4 file format through a ring buffer supporting concurrent read and write operations.
To meet subsequent analysis processing requirements, the system stores event data and associated metadata in a custom data format. 
Furthermore, by embedding a data decoding module, it achieves efficient integration with the REST framework, ultimately completing the unified integration of data processing and analysis functions.

\subsection{Communication with the electronics}

The PandaX-III experiment integrates two types of backend electronics, each employing dedicated communication methods to fulfil its requirements.
 
The DCM backend electronics utilize a USB 3.0 interface to achieve high-speed data transmission between the server and front-end electronics, with a theoretical bandwidth of up to 1.6~GB/s. 
The communication protocol is built on the CYUSB driver framework, ensuring cross-platform compatibility and low latency.
After sending a configuration file via the CYUSB driver to initialize the DCM back-end electronics, the interface displays the information, including the bulk transaction mode and input/output ports.
The USB transfer packet size is 4096 bytes, and the data read buffer is also configured with 4096 bytes as a unit to facilitate decoding.

On the other hand, the TDCM backend electronics employ the User Datagram Protocol~(UDP) over Ethernet via network cables for data transmission. 
This protocol is selected for its low-latency and connectionless characteristics, making it well-suited for scenarios requiring continuous, real-time data streaming. 
The TDCM transmission relies on standard network drivers, ensuring broad compatibility with mainstream server operating systems and simplifying system integration into the overall DAQ workflow.

\subsection{Web interface}

\begin{figure}[htbp]
\centering
\includegraphics[width=\hsize,trim=0 0 0 0,clip]{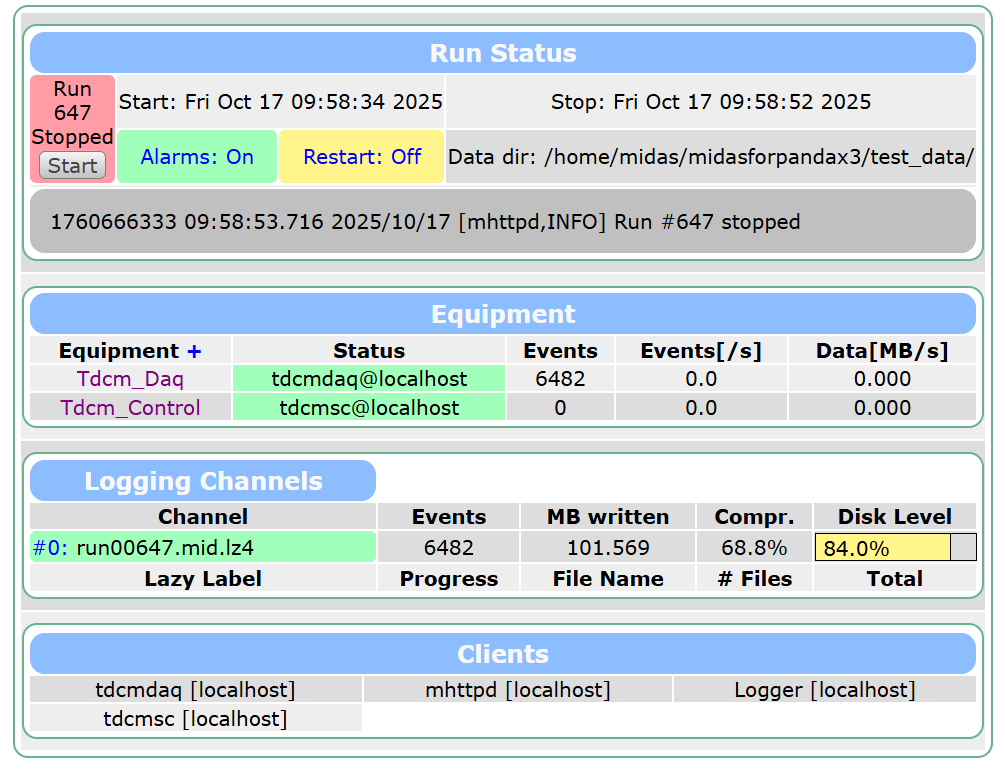}
\caption{The main interface of the DAQ web page includes the data acquisition button, acquisition time, data storage path, and event rate.}
\label{DAQ_interface}
\end{figure}

As a core user interaction portal for the DAQ system, our DAQ web interface is modularly reorganized based on MIDAS’ native capabilities, tailored to the needs of the PandaX-III experiment. 
As shown in Fig.~\ref{DAQ_interface}, it features four core interaction areas. 

\begin{itemize}
    \item The first part of the web interface is the \enquote{Run Status}, which serves as the central control hub for experiment operation. 
    It displays real-time critical information, including experiment start/end times, system alarm notifications, and data storage paths. 
    It also provides four core run control commands~(Start, Stop, Pause, Resume) via intuitive buttons, enabling users to adjust the experiment’s operational state directly.
    \item The second part, \enquote{Equipment}, visualizes key metrics: the operational status of the defined device, the number of collected events, as well as the event rate and data transfer rate.
    \item The third part, \enquote{Logging Channels}, focuses on data storage information: the number of events read into the file and the disk usage.
    \item The fourth part, \enquote{Clients}, lists all running MIDAS-based applications that underpin both the \enquote{CmdProc} and \enquote{DataFlow} links.
    These contain: tdcmdaq, tdcmsc, mhttpd, and Logger. 
    The tdcmsc program contains rules for sending configuration commands~(CmdProc link).
    And the tdcmdaq program implements the rules for reading raw data, building events, and temporary data caching~(DataFlow link).
    The mhttpd program hosts the web interface itself. 
    To ensure secure access, users can forward the server’s web port to their local machine via SSH, enabling them to open and operate the interface through a local browser without direct public network access.
    The Logger program is MIDAS’ built-in data storage tool that operates under the pre-specified file path. 
\end{itemize}

\subsection{Data format}

\begin{figure}[htbp]
\centering
\includegraphics[width=\hsize,trim=80 80 65 70,clip]{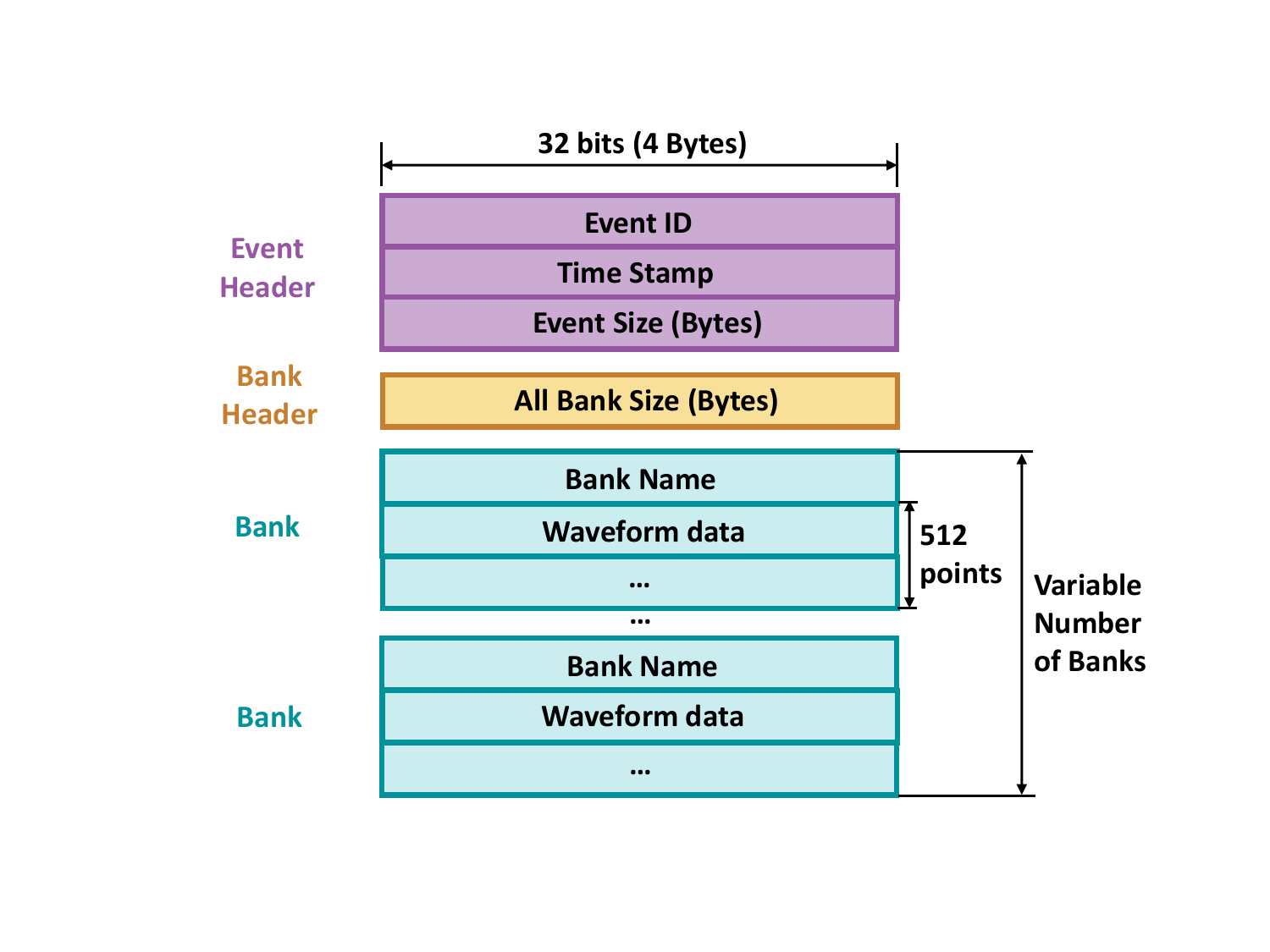}
\caption{The simplified data format of the DAQ.}
\label{Data_format}
\end{figure}

After parsing the received data frame, the unpacked waveform data from the same event will be packed following the data format, which is a hierarchical event structure design of Midas, as shown in Fig.~\ref{Data_format}. 
Event data comprises three components: the event header, the global bank header, and the data bank, where each bank contains the data for a single channel. 
Such a standardized format ensures data integrity and traceability.

The purple part in Fig.~\ref{Data_format} is the event header, which contains three units~(each unit is four bytes). 
The first unit, \enquote{Event ID}, is an event counter that increments from 0. 
The \enquote{Time Stamp} represents the receiving time of the event by DAQ.
The Event Size refers to the byte size of the current event, excluding the event header.
The yellow section denotes the global bank header, which consists of one unit: \enquote{All Bank Size}. 
\enquote{All Bank Size} indicates the total byte size of all banks, excluding the global bank header itself. 
Waveform sampling data is stored in the blue data banks. 
The Bank Name~(4 bytes) displays encoding physical location information: the first two bytes are the fibre port number~(range: 1-32), the third byte is the AGET chip number~(range: 1-4), and the last byte is the channel number~(range: 1-68).
For example, bank name 1019 indicates this bank stores data from channel 9 of the first AGET chip in the FEC connected to fiber optic port 10.
Each channel is fixed at 512 samples, so the data portion of the bank totals 1024 bytes. 
The total byte count for reading out all 6,656 channels of the detector is 6.5~MB, representing the maximum size of a single event.

\subsection{Analytical Framework}
 
The DAQ software is fully interfaced with the data analytics framework, breaking through the limitations of traditional data collection and significantly reducing the time required to transform raw data into actionable insights.
The DAQ system features real-time monitoring capabilities, enabling simultaneous display of signal waveforms and energy spectrum distributions during the data acquisition process. 
This allows users to assess signal quality in real time, dynamically adjust experimental parameters, and quickly identify optimal configuration settings, which significantly enhances the controllability and flexibility of the experimental process.

Among these, the integration of the REST framework~(Rare Event Searches Toolkit) serves as the core technological foundation for the DAQ software's extended functionality. 
As a general-purpose toolkit specifically designed for gaseous TPC experiments, REST is developed using the C++ language and establishes a complete, traceable chain for processing complex events. 
Through three interfaces—event type, metadata, and event stream—it forms a complete processing chain from raw signal acquisition, waveform analysis, feature extraction, to trajectory reconstruction. 
This design ensures high traceability in the data processing process, with parameters and results of each intermediate processing step fully recorded, providing a detailed basis for subsequent data verification and algorithm optimization.

In practical applications, researchers only need to define the hierarchical structure and key parameters of the data processing chain in an XML-formatted configuration file to complete the customization of the analysis workflow. 
The configuration file encapsulates all process information, including data sources, processing algorithms, and output formats, ultimately generating standardized ROOT-format output files. 
This file not only stores all variables generated during the analysis process but also supports flexible retrieval and in-depth exploration of data through REST metadata objects. 
Through this modular, parameterized design, the DAQ software achieves rapid adaptation to different experimental scenarios, effectively enhancing the system's versatility and scalability.

\section{Functional Overview}\label{sec.IV}

The DAQ software enables the electronic parameter configuration via the \enquote{cmdProc} application and ODB in the MIDAS web interface.
The parameter configuration.
To balance efficiency and flexibility, a dual-mode configuration mechanism has been designed through the synergistic use of XML global configuration and ODB hot-linking mechanisms.

\begin{figure}[htbp]
\centering
\includegraphics[width=\hsize, height=0.9\textheight, keepaspectratio, trim=130 20 130 20,clip]{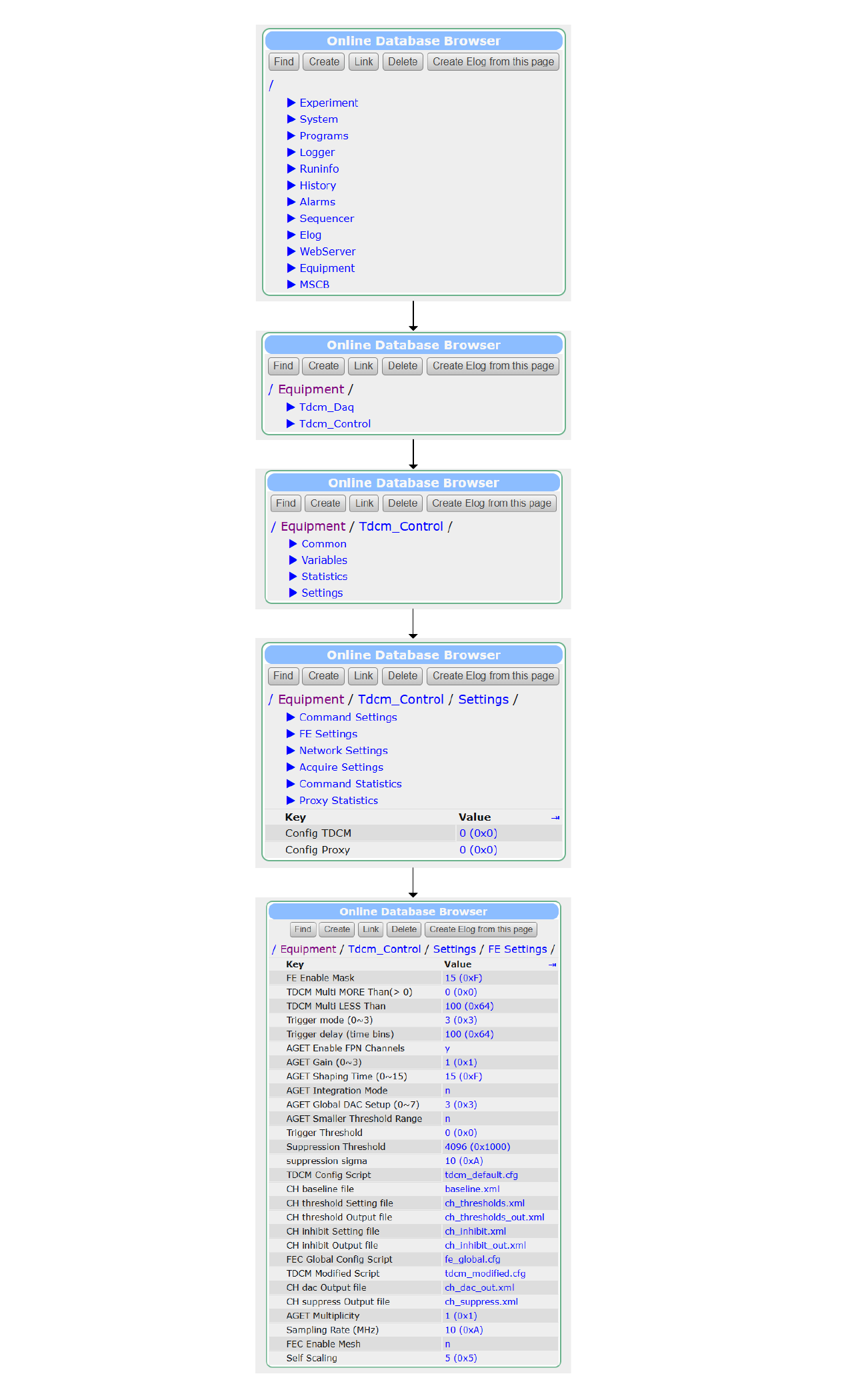}
\caption{The flow follows the hierarchical path: /Equipment /Tdcm\_Control /Settings /FE Settings in the Online Database Browser, presenting the core key-value parameter configuration for electronic components. One-click configuration is performed via the \enquote{Config TDCM} key in the \enquote{Settings} directory.}
\label{para_config}
\end{figure}

The one-click global configuration via XML files is suitable for repetitive experiments and standardized testing.
Based on the ODB hot-linking mechanism, this software enables real-time dynamic parameter modification, suitable for fine-tuning individual parameters during experiments. 
Fig.~\ref{para_config} illustrates the process of parameter configuration using the hotlink mechanism.
The subdirectories of the device ODB are organised in a tree structure, with each node corresponding to different electronic parameters in key-value format. 
Users may modify the parameter configuration of individual variables by altering the key-value pairs in the ODB database; for example, set the \enquote{Config TDCM} key-value in Fig.~\ref{para_config} to 0 to issue commands.

The BEC provides 32 fiber optic ports that communicate with the FECs, with key value \enquote{FE Enable Mask} being used to configure the port number for our communication.

Three trigger modes have been implemented to meet different experimental objectives, with the key \enquote{Trigger mode} used to configure the AGET trigger mode in the ODB, which includes:
\begin{itemize}
    \item Self-Trigger: Triggered internally by the FPGA on the FEC, with a trigger frequency of 24~Hz. This mode is primarily used for testing the baseline and noise performance of AGET chips.
    \item Hit Trigger: Triggered by the AGET chips directly, suitable for general physical data acquisition.
    \item Multiplicity Trigger: AGET chips operate in multiplicity mode, where the trigger is generated based on the set number of channels that detect hits and send signals to the TDCM.
\end{itemize}

The amplitude of an electrical signal waveform corresponds to the energy of a single channel, with a dynamic range of 4096 ADC counts. 
To avoid low-energy signals drowning in noise and high-energy signals overthresholding, the AGET chip provides a 4-step dynamic range of 120~fc, 240~fc, 1~pc, and 10~pc to correspond to the range to ensure that the waveform amplitude is appropriate. 
The values 0,1,2, and 3 of \enquote{AGET Gain} in the online database correspond to the dynamic ranges 120~fc, 240~fc, 1~pc, and 10~pc, respectively.

When one or several channels are triggered, all channels in the entire readout plane will be read out, which will take up a large amount of data transmission bandwidth. 
The AGET chip has a channel compression function to remove other untriggered channel data~\cite{}.
Users can configure the channel compression mode via the \enquote{Suppression Threshold} key value in the ODB. 
\begin{itemize}
    \item The key value 0 represents no channel compression and is usually selected when acquiring a noise baseline.
    \item The key value from 1 to 4095 represents a fixed value for compression from all channels, under the condition that the single channel range is 4096~ADC.
    \item The key value 4096 signifies that the channel compression threshold is calculated as the mean value per channel plus a certain multiple of the sigma value.
    And a certain multiple of the sigma value is configured within the \enquote{suppression sigma}.
    Different channels have different compression thresholds, and the channel compression threshold setting increases as the noise baseline increases. 
    Noise acquisition can be performed prior to each data acquisition run, which in turn automatically generates a compression threshold profile based on the mean and variance of the noise for each channel.
\end{itemize}

The AGET chip contains two fixed ranges of trigger thresholds: 5\% and 17.5\% of the dynamic range, each divided into 8 equalized steps to allow for a more accurate selection of the trigger thresholds. 
The n for the \enquote{AGET Smaller Threshold Range} key value in ODB means that 17.5\% of the dynamic range is selected as the total trigger threshold, and y means that 5\% is chosen as the total trigger threshold. 
The \enquote{AGET Global DAC Setup~(0-7)} in the ODB has a total of eight integers from 0-7 for the eight slots, divided based on the total trigger threshold. 
For example, the \enquote{AGET Gain~(0-3)} key value of 1 in the figure represents a dynamic range of 240~fc; \enquote{AGET Smaller Threshold Range} is n and \enquote{AGET Global DAC Setup~(0-7)} is 3, which represents a trigger threshold of 17.5\% of the dynamic range multiplied by a factor of four-eighths, i.e., 21~fc.

The AGET chip has 512 sampling points, with the key value of  \enquote{Trigger delay} determining where the waveform triggers.
And the sampling rate can be selected from 5~MHz to 100~MHz, which can be chosen according to different physical events. 
The trigger delay time can also be adjusted to set the trigger position of the signal in the time window so that the complete waveform can be displayed according to the time window size. 
Events with long trajectories and long trigger times, such as muons, require long windows with low sampling rates, whereas data acquisition for some radioactive sources can employ short time windows with high sampling rates.
The key value of \enquote{Sample Rate~(MHz)} in the ODB can modify the sample rate size setting, the value of 10 represents the setting of 10~MHz sampling rate, and the time window is 51.2~\textmu s.

\section{Performance test}\label{sec.V}

To validate the reliability and adaptability of the developed universal DAQ software for the PandaX-III experiment, we have performed basic electronic functional tests using a signal generator, as well as systematic joint tests with the aforementioned FEC and BEC modules.
The integrated data acquisition system can be successfully applied to the PandaX-III prototype testing phase. 
The following subsections present detailed performance test results, focusing on verifying the software’s functional integrity, data transmission stability, and compatibility with detector readout modules under different experimental configurations. 

\subsection{Experimental setup}\label{sec.IV}

\begin{figure}[htbp]
    \centering   
    \subfigure[]{
        \label{detector}
        \includegraphics[width=.42\hsize,height=6cm]{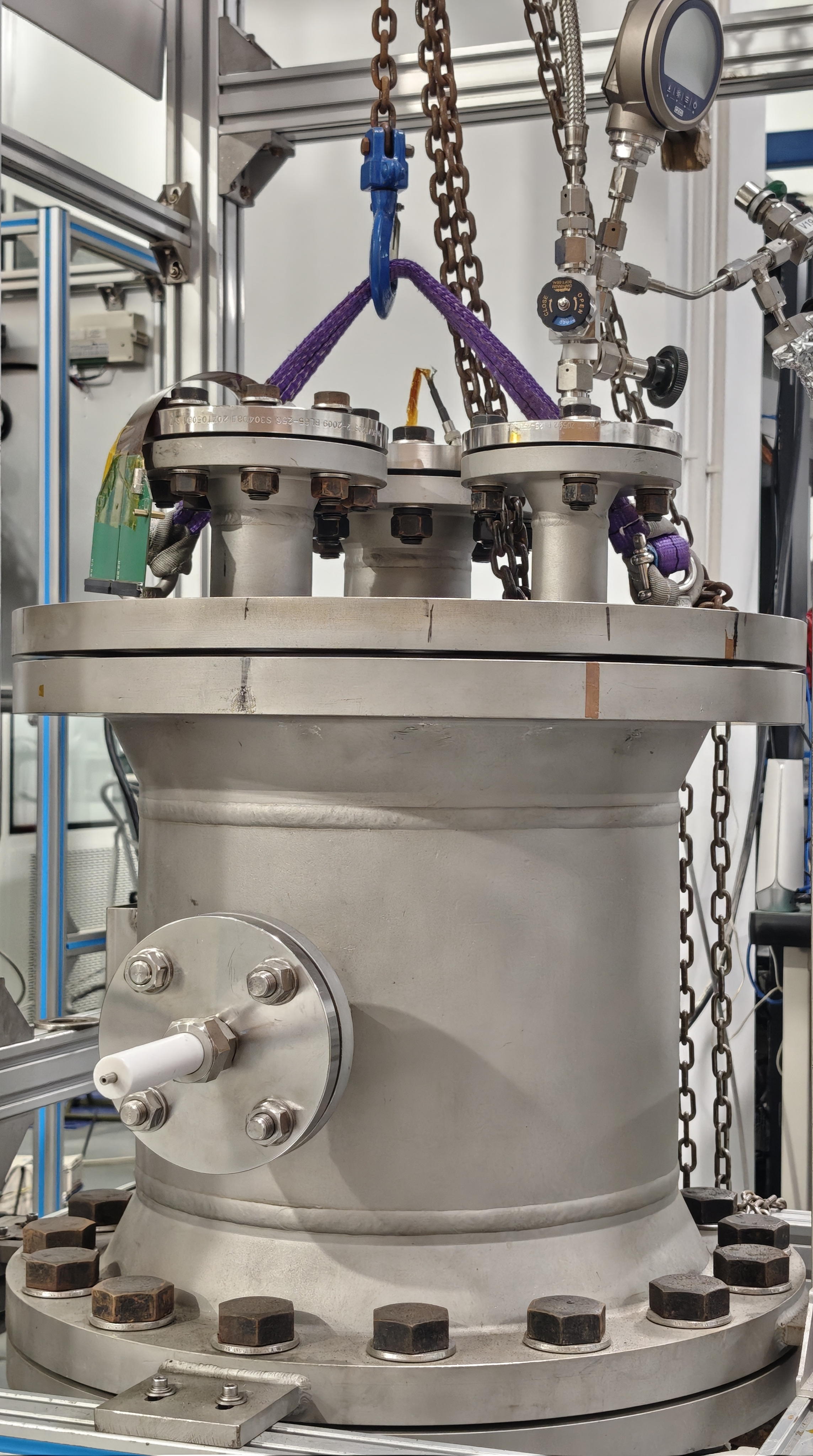}
    }    
    \subfigure[]{
        \label{micromegas}
        \includegraphics[width=.42\hsize,height=6cm]{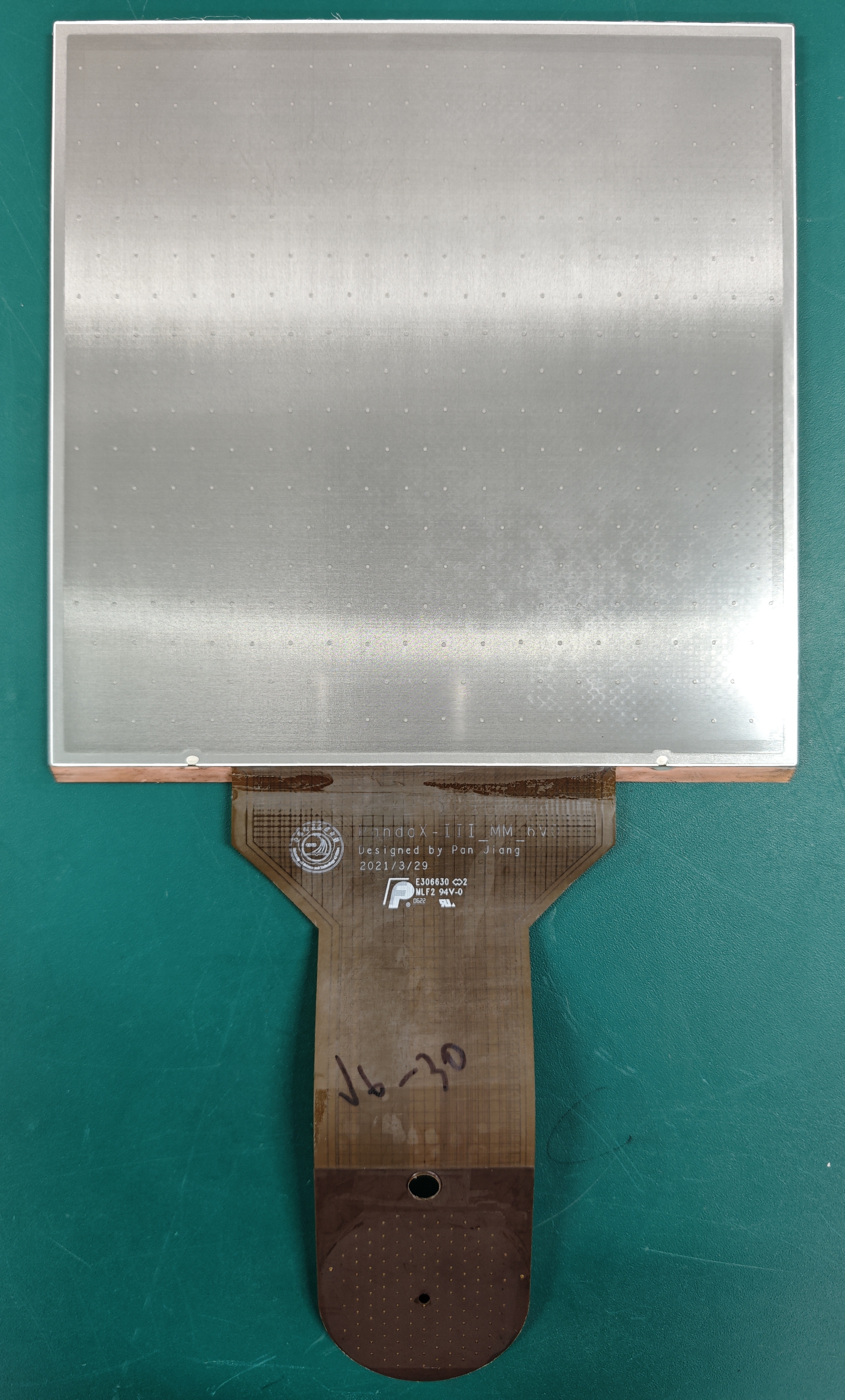}
    }
    \caption{\label{experiment setup}
    Detector vessel and readout module.
    (a) A stainless steel vessel for the gaseous TPC detector. 
    The side flange accommodates a high-voltage feedthrough, while the top flange connects to the front-end electronics adapter board.
    (b) Picture of a $20 × 20\ \text{cm}^2$ thermal bonding Micromegas.}
\end{figure}

For performance validation of the complete data acquisition system, a gaseous TPC with a 20~cm drift distance is employed, as depicted in Fig.~\ref{detector}.
The core components of the detector include a bottom cathode, a middle field cage for maintaining a uniform electric field, and a top charge readout plane. 
The sensitive volume of the detector is approximately 8~L.
The field cage, constructed from acrylic and copper rings, has a diameter of 34~cm and a height of 20~cm. 
The readout plane is a 20~cm-sided Micromegas module developed by the University of Science and Technology of China, as shown in Fig.~\ref{micromegas}.
The module features a total of 128 readout channels, equally divided into 64 channels along the X and 64 along the Y direction, with a pitch of 3~mm.
The working gas is argon mixed with isobutane.

Two calibration sources, designated as gaseous \ce{^{37}Ar} and solid \ce{^{109}Cd}, were employed to evaluate system performance.
\ce{^{37}Ar} decays via electron capture to \ce{^{37}Cl}~(half-life: 35 days), emitting X-rays with an energy of ~2.82 keV. 
\ce{^{109}Cd} decays via electron capture to \ce{^{109}Ag}~(half-life: 463 days), primarily emitting 22~keV X-rays.
During the experiment, long-term measurements were conducted on two radioactive sources under different drift fields and gas pressure conditions to comprehensively evaluate the performance of the detector and DAQ software working together.

\subsection{Basic electronic test}

\begin{figure}[htbp]
\centering
\includegraphics[width=\hsize]{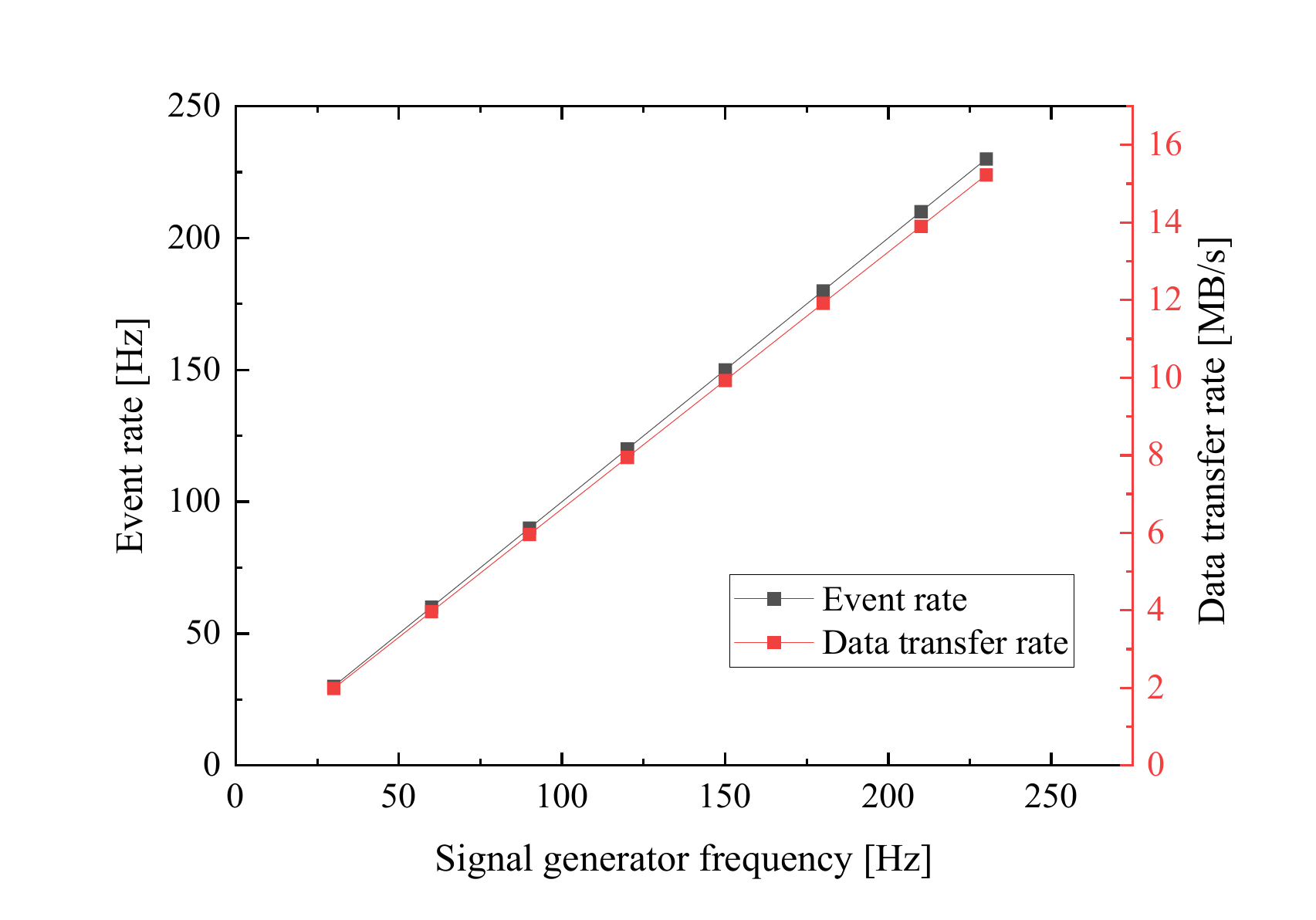}
\caption{The relationship between the event rate, data transfer rate, and signal generator frequency under full readout conditions for a single AGET chip.}
\label{SingleDataTransfer}
\end{figure}

Accuracy and stability of data transmission are the key performance indicators of DAQ.
We use a signal generator to generate signals of different frequencies as inputs. 
For full readout of 64 channels in a single AGET chip, the event rate and data transfer rate observed by the software are shown in Fig.~\ref{SingleDataTransfer}.
The event rate is basically the same as the signal generation frequency when the setting frequency is lower than 230~Hz, where the rate of a single AGET full readout is 15.2~MB/s.
Long-term operational stability was evaluated over a cumulative period of one month, comprising many consecutive over-10-hour data acquisition runs with intermittent interruptions for hardware configuration adjustments and gas system operations. 
The rate of data packet loss and data corruption during operation was negligible, fully demonstrating the exceptional reliability and robustness of this data acquisition system during long-term experimental operation.

\begin{figure*}[htbp]
    \centering   
    \subfigure[]{
        \label{SamplingRate5M}        \includegraphics[width=0.47\textwidth,trim=10 15 35 30,clip]{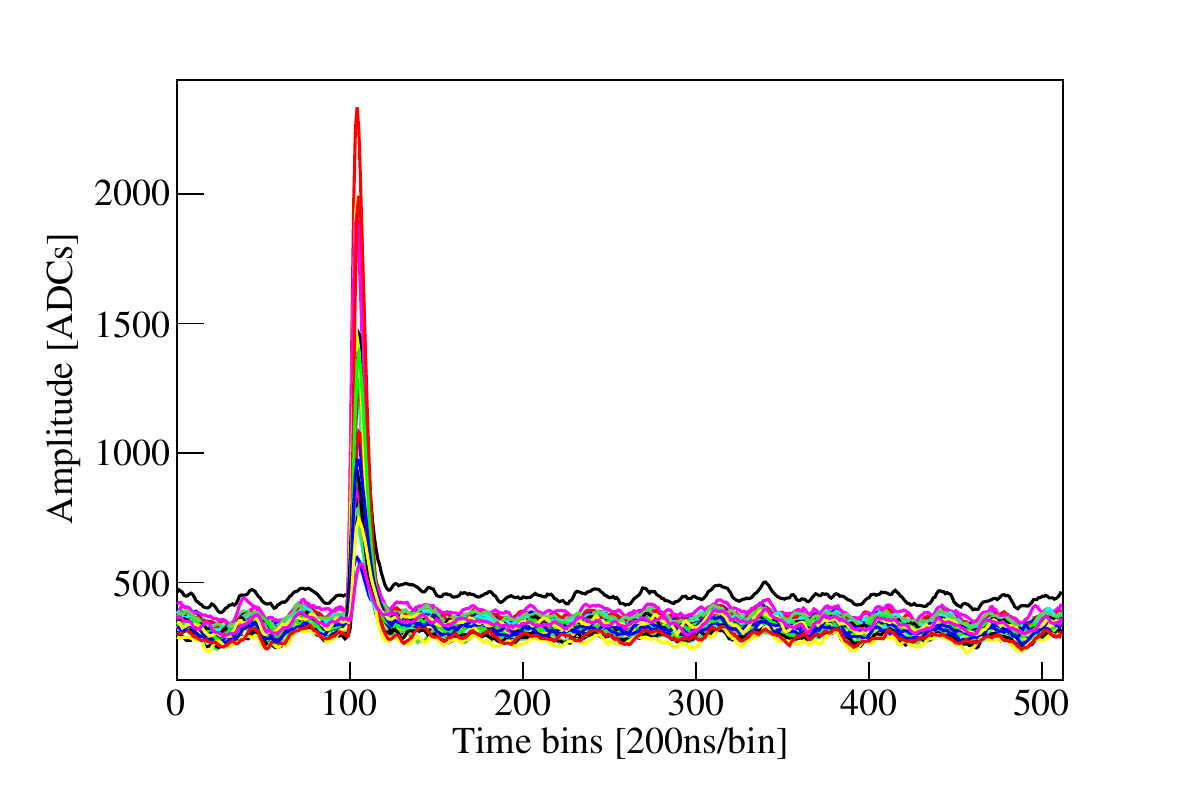}
    }    
    \subfigure[]{
        \label{SamplingRate100M}        \includegraphics[width=0.47\textwidth,trim=10 15 35 30,clip]{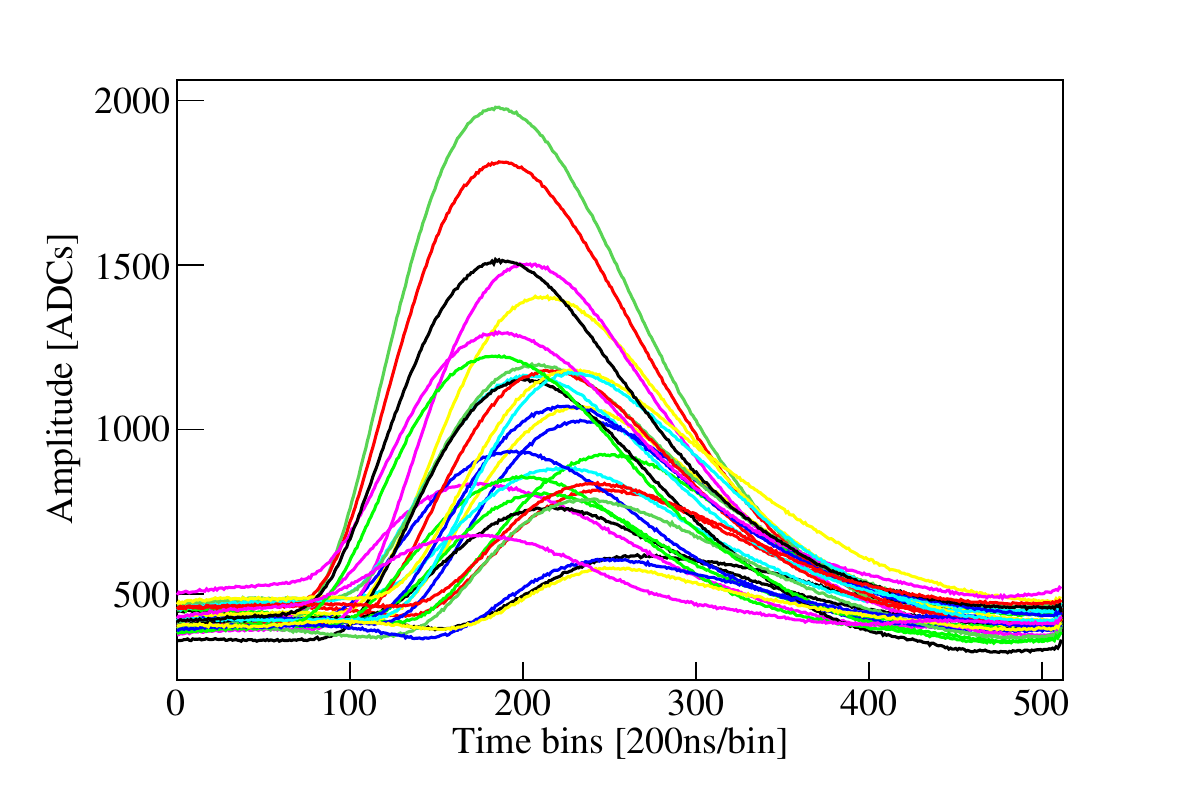}
    }    
    \subfigure[]{
        \label{TriggerDelay200}        \includegraphics[width=0.47\textwidth,trim=10 15 35 30,clip]{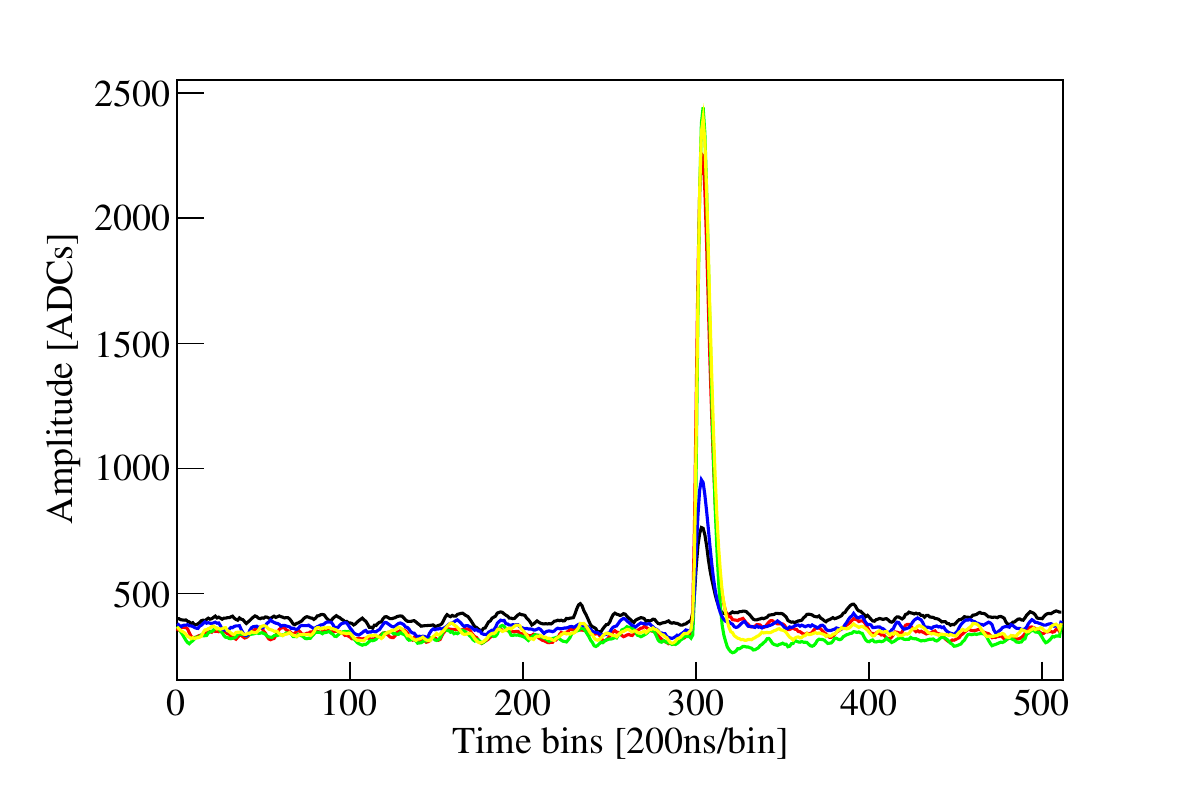}
    }    
    \subfigure[]{
        \label{NoCompression}        \includegraphics[width=0.47\textwidth,trim=10 15 35 30,clip]{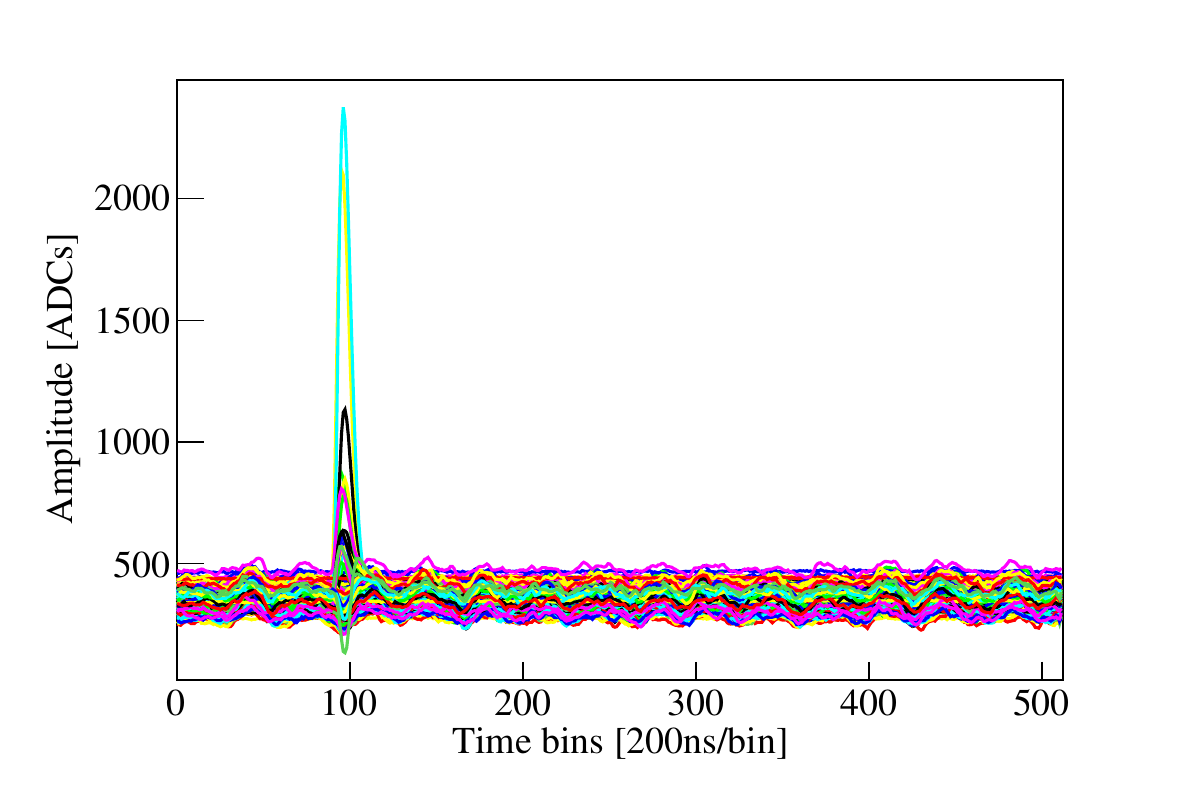}
    }    
    \subfigure[]{
        \label{noise}        \includegraphics[width=0.47\textwidth,trim=10 15 35 30,clip]{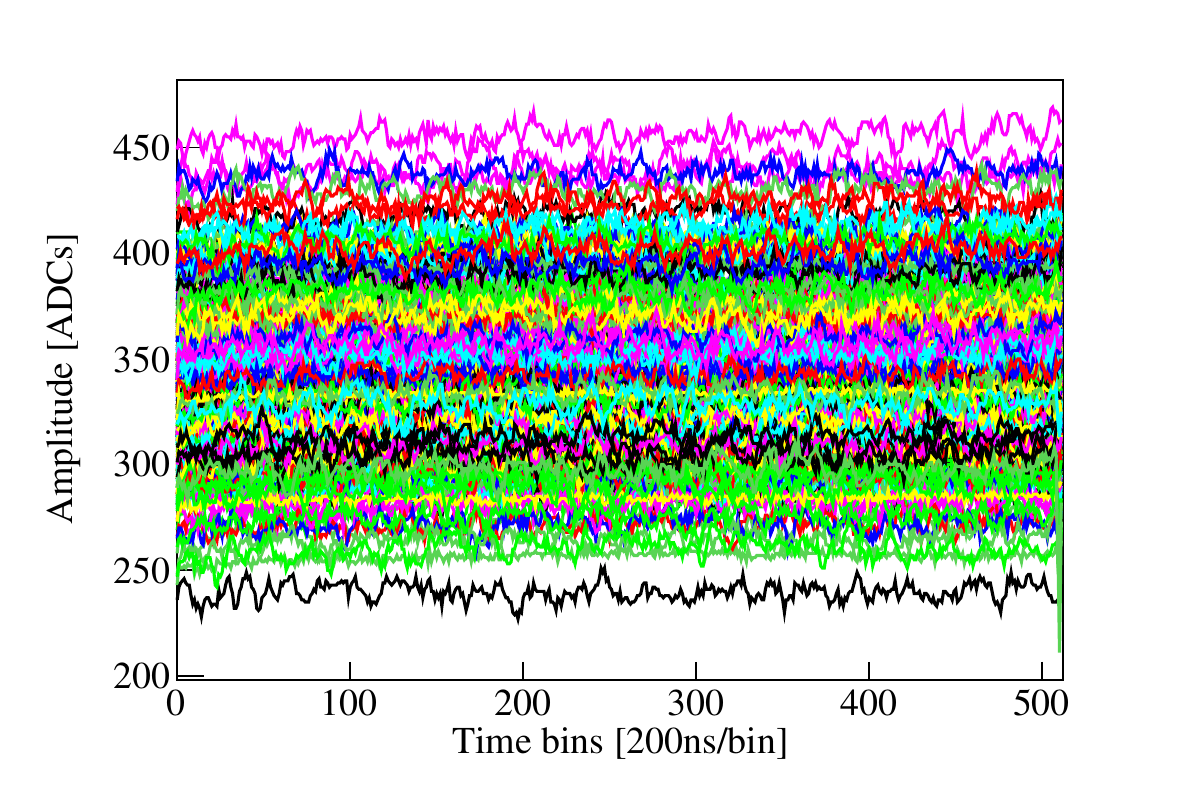}
    }    
    \subfigure[]{
        \label{self-calibration}        \includegraphics[width=0.47\textwidth,trim=10 15 35 30,clip]{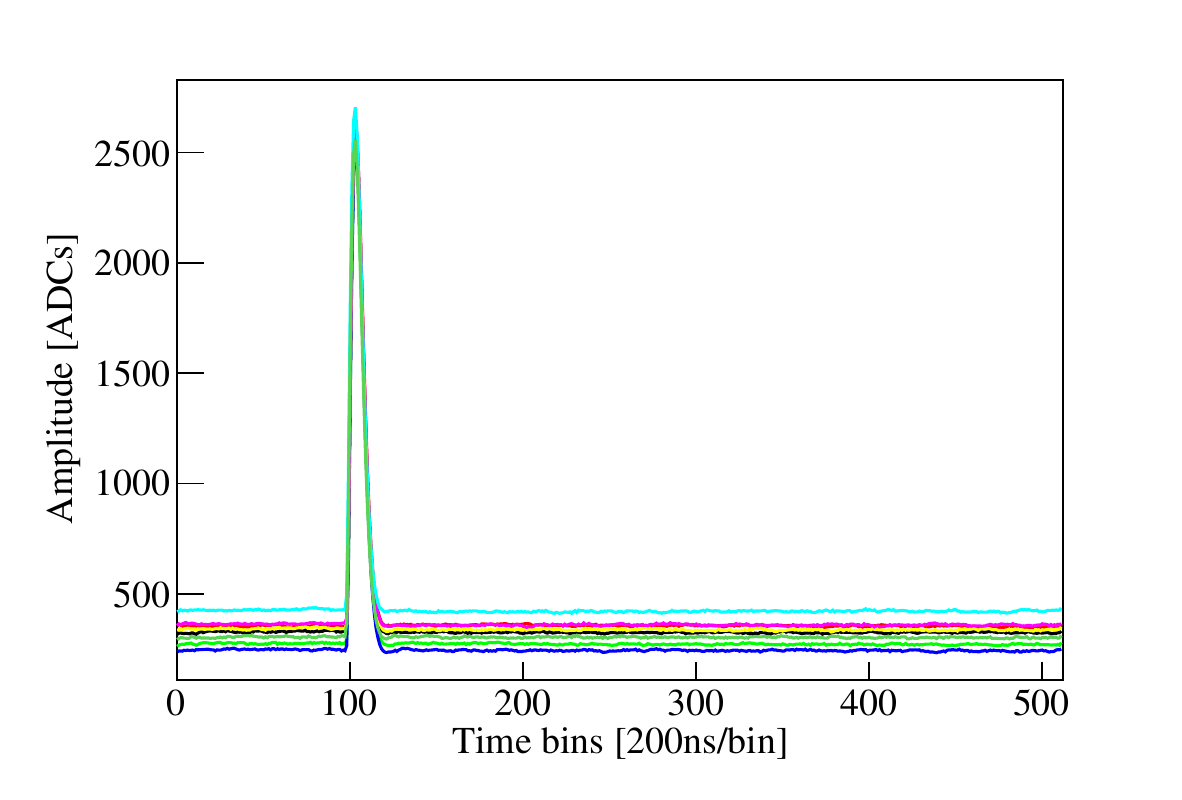}
    }    
   
\caption{\label{DelayandChannelCompressionandSamplerate}
    Physical signals under different parameters, noise waveforms, and self-calibrated signals.
    (a) \ce{^{37}Ar} signals under the conditions of channel compression, a trigger delay of 100 points, and a sampling rate of 5~MHz.
    (b) \ce{^{37}Ar} signals under the conditions of channel compression, a trigger delay of 100 points, and a sampling rate of 100~MHz.
    (c) \ce{^{37}Ar} signals under the conditions of channel compression, a trigger delay of 300 points, and a sampling rate of 5~MHz.
    (d) \ce{^{37}Ar} signals under the conditions of full-channel readout, a trigger delay of 100 points, and a sampling rate of 5~MHz.
    (e) Electronics noise waveforms.
    (f) Electronics self-calibrating signals.}
\end{figure*}

The implementation of core electronic functions of the DAQ software constitutes the fundamental basis for data collection, providing a critical prerequisite for subsequent joint detector commissioning.
This test focuses on its performance in processing physical signals, including waveform acquisition, sampling rate adjustment, trigger delay control, channel compression, noise characterization, and electronic self-calibration.
The test was conducted using a single readout module and its matching small detector, with the hardware configuration illustrated in Fig.~\ref{experiment setup}. 
Fig.~\ref{DelayandChannelCompressionandSamplerate} shows the waveforms of the \ce{^{37}Ar} source at an amplified field of 53~kV/cm with different sample rates, trigger delays, and channel compression in a 3~bar gas mixture of argon with 2.5\% isobutane.

 Fig.~\ref{SamplingRate5M} displays a typical \ce{^{37}Ar} signal waveform, which serves as a reference benchmark for comparing other test results. 
 This waveform was acquired under the following test conditions: 5~MHz sampling rate, 100 points trigger delay, and enabled channel compression. 
 Fig.~\ref{SamplingRate100M} shows the \ce{^{37}Ar} waveform obtained at a higher sampling rate of 100~MHz,  maintaining the same trigger delay and channel compression settings, indicating that the DAQ software can accommodate different sampling frequency requirements. 
 Fig.~\ref{TriggerDelay200} presents the waveform when the trigger delay is set to 300 points, demonstrating that the number of trigger delay points can be flexibly set according to the characteristics of the signal waveform. 
 Fig.~\ref{NoCompression} exhibits the full \ce{^{37}Ar} source waveform without channel compression, where it can be seen that most channels are noise data triggered by non-signals, which will occupy a certain amount of bandwidth. 
 Additionally, Fig.~\ref{noise} records the electronic noise level of the system, while Fig.~\ref{self-calibration} displays the signal generated during the electronic self-calibration process.
 The above results are comparative diagrams of the various test functions, all of which validate the reliability of the data acquisition system’s basic functions.
    
Collectively, these waveform results confirm that all targeted basic electronic functions of the data acquisition software have been successfully implemented.

\subsection{Joint test with the detector}\label{singleModule}

\begin{figure*}[htbp]
    \centering   
    \subfigure[]{
        \label{Ar-37map}
        \includegraphics[width=0.42\textwidth,trim=50 60 30 100,clip]{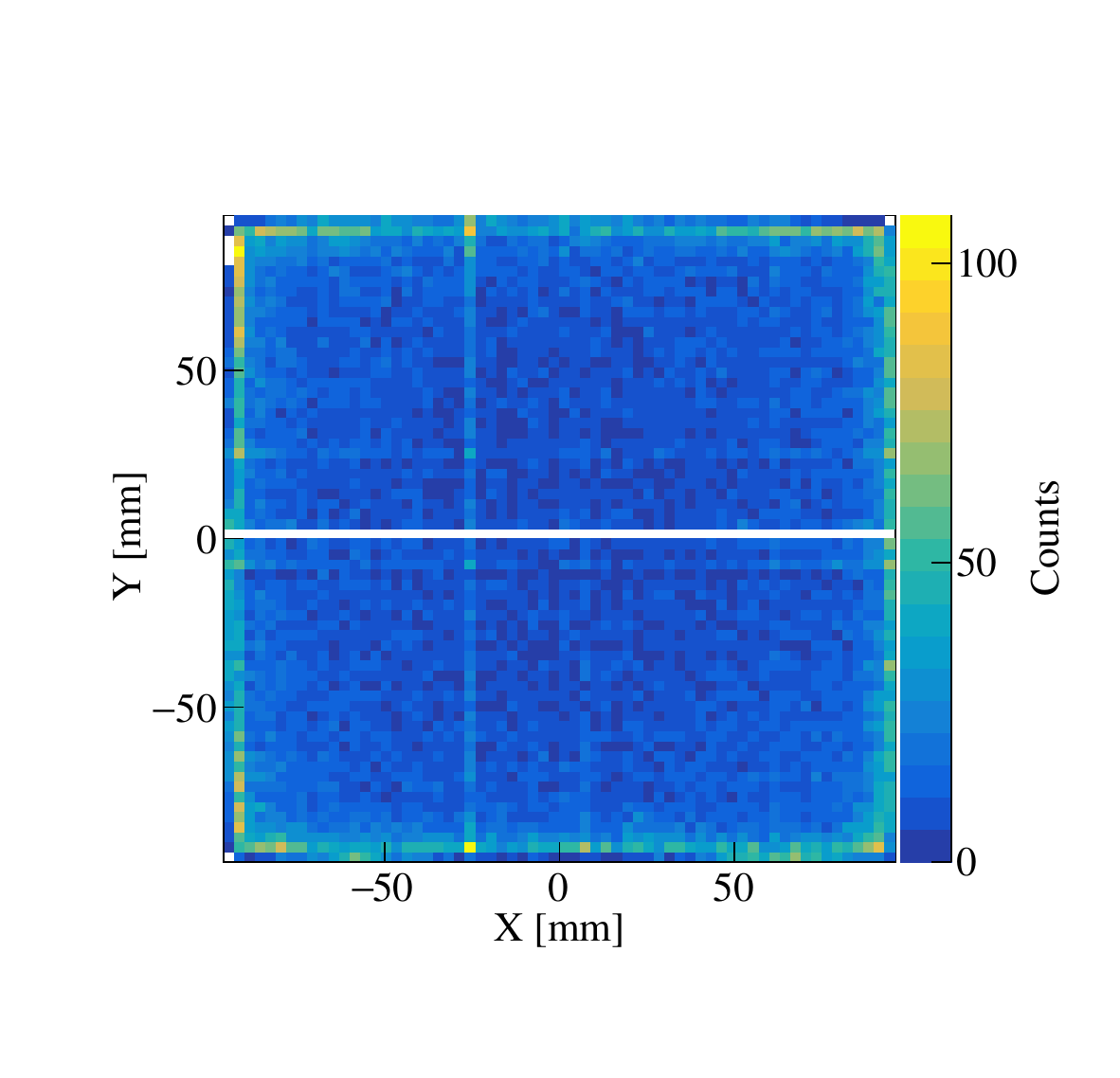}
    }    
    \subfigure[]{
        \label{Cd-109map}
        \includegraphics[width=0.42\textwidth,trim=50 60 30 100,clip]{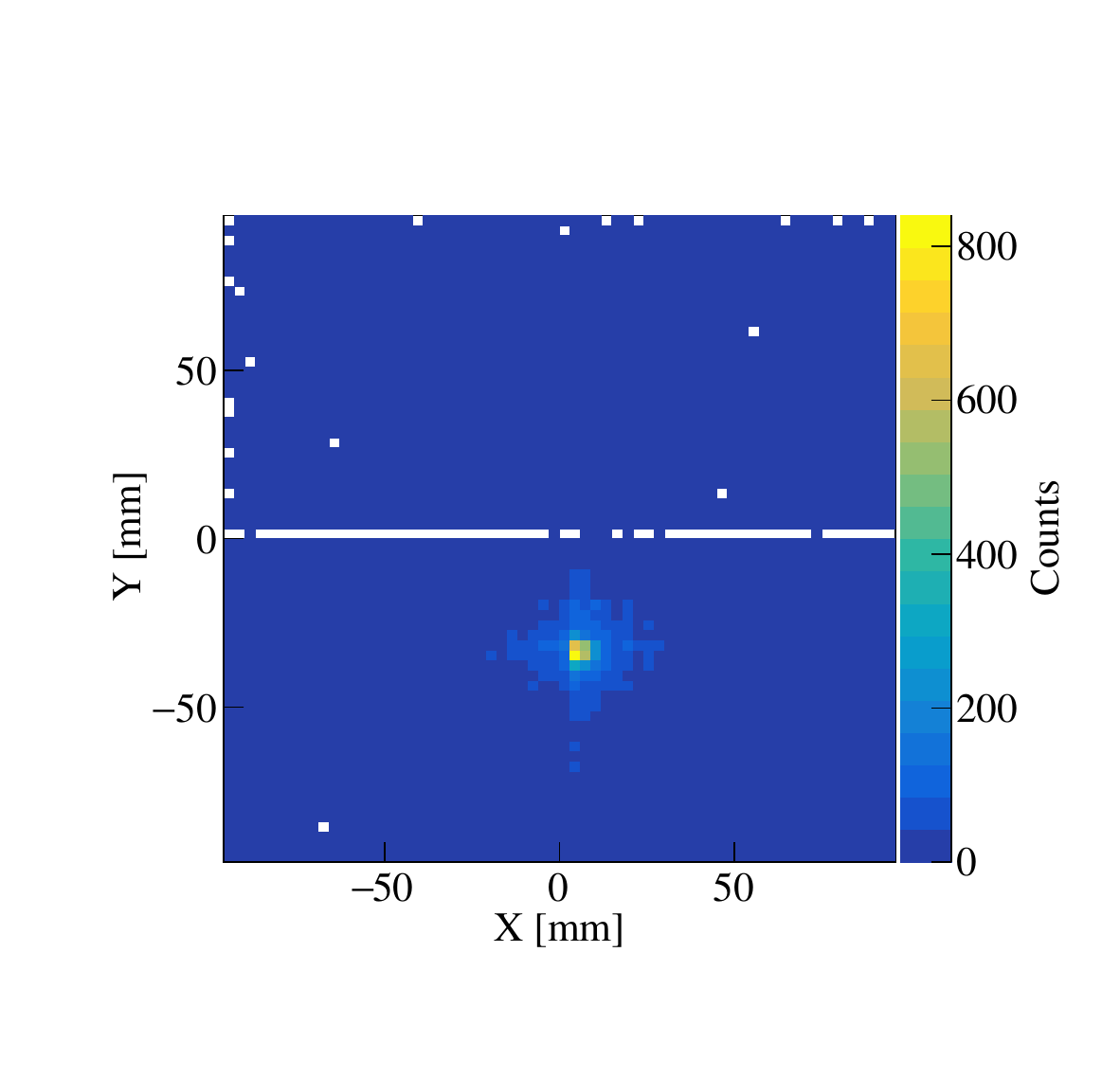}
    }
    \subfigure[]{
        \label{Ar-37energy2}
        \includegraphics[width=0.42\textwidth,trim=50 60 35 100,clip]{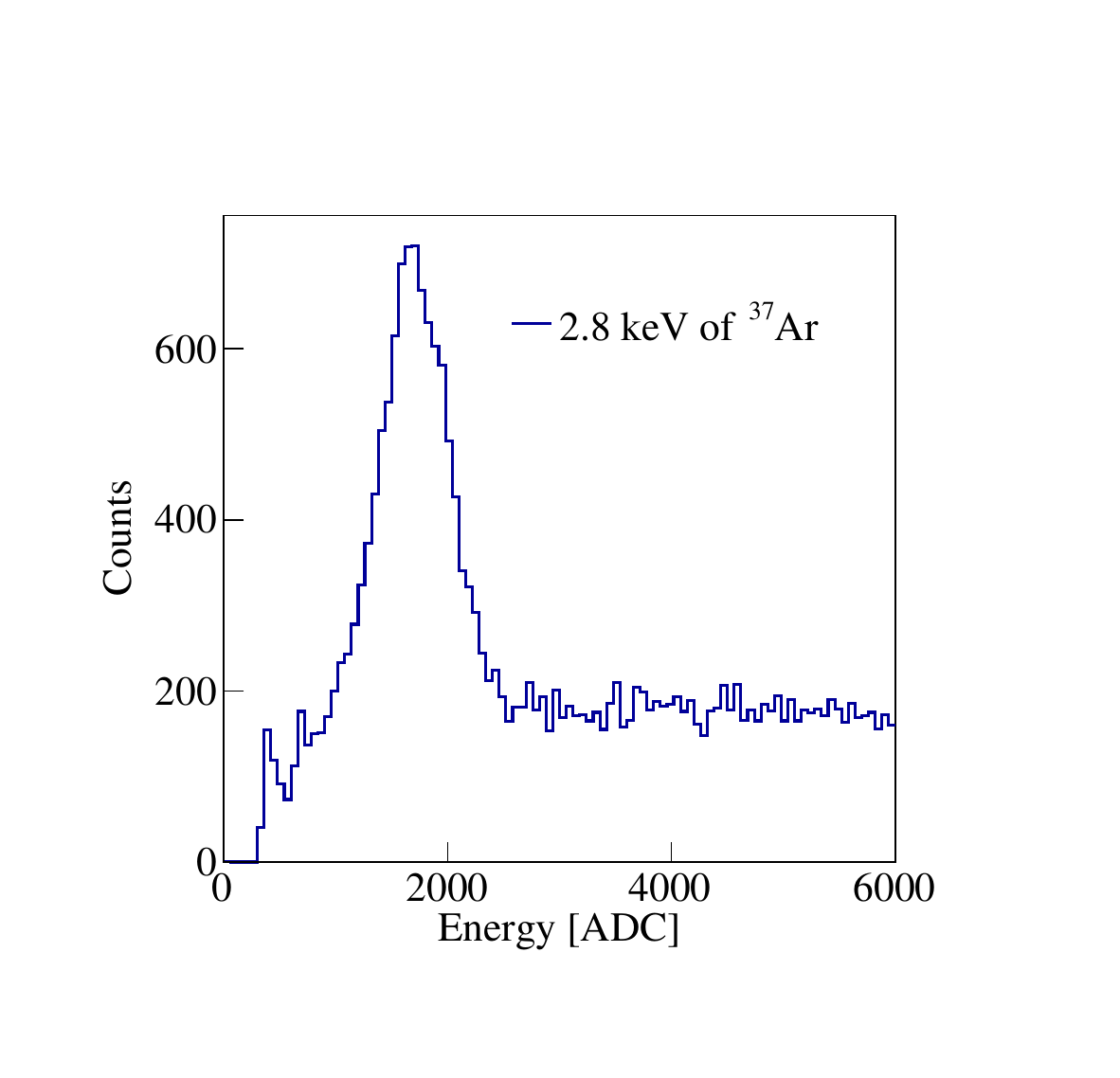}
    }
    \subfigure[]{
        \label{Cd-109energy2}
        \includegraphics[width=0.42\textwidth,trim=50 60 35 100,clip]{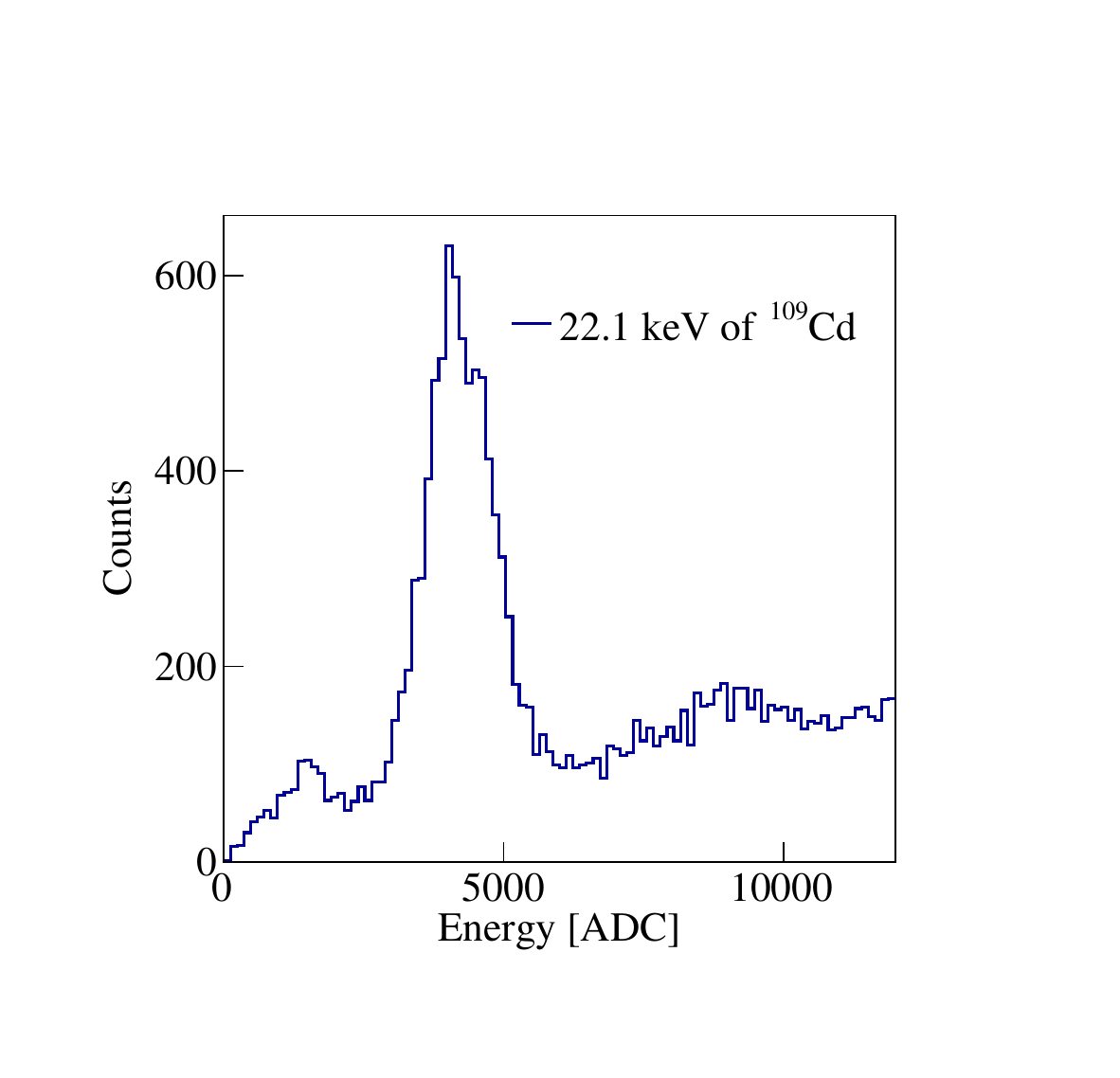}
    }
    \caption{\label{hitmap}
    Hitmap and energy spectrum from the single readout module test.
    (a) A hit position map of \ce{^{37}Ar} source tested with the single readout module.
    The white area where Y=0 is the electronic bad channel.
    (b) A hit position map of X-rays from \ce{^{109}Cd} source tested with the single readout module.
    The bright spot indicates the position of the \ce{^{109}Cd} source on the XY plane.
    (c) A energy spectrum of \ce{^{37}Ar} source.
    (d) A energy spectrum of X-rays from \ce{^{109}Cd} source.}
\end{figure*}

Building upon the successful validation of the DAQ software's fundamental functionalities, this section presents the results of integrated tests conducted with a single readout module. 
As depicted in Fig.~\ref{hitmap}, the DAQ system was employed to map the two-dimensional (XY) signal distributions and corresponding energy spectra of the gaseous \ce{^{37}Ar} source and solid \ce{^{109}Cd} source on the detector readout plane.
Specifically, Fig.~\ref{Ar-37map} and Fig.~\ref{Ar-37energy2} show the hitmap and energy spectrum of the \ce{^{37}Ar} source, measured at a dynamic range of 240~fc under the following conditions: 3~bar argon mixed with 2.5\% isobutane, a drift field of 348.5~V/cm, and an amplification field of 53~kV/cm. 
The observed concentration of event clusters at the edges of the readout plane is attributed to electric field distortions, a phenomenon consistent with previous work~\cite{b27}.
Similarly, Fig.~\ref{Cd-109map} and Fig.~\ref{Cd-109energy2} present the hitmap and energy spectrum of the \ce{^{109}Cd} source. 
These measurements were carried out in 10~bar argon mixed with 2.5\% isobutane, featuring a drift field of 550~V/cm, an amplification field of 110~kV/cm, and a dynamic range of 1~pc. The blank region observed along the Y=0 channel in the hitmap corresponds to electronic bad channels, as indicated in the figure caption.

\begin{figure*}[htbp]
    \centering
    \subfigure[]{
        \label{readout_plane}        \includegraphics[width=0.42\textwidth,trim=0 0 0 0,clip]{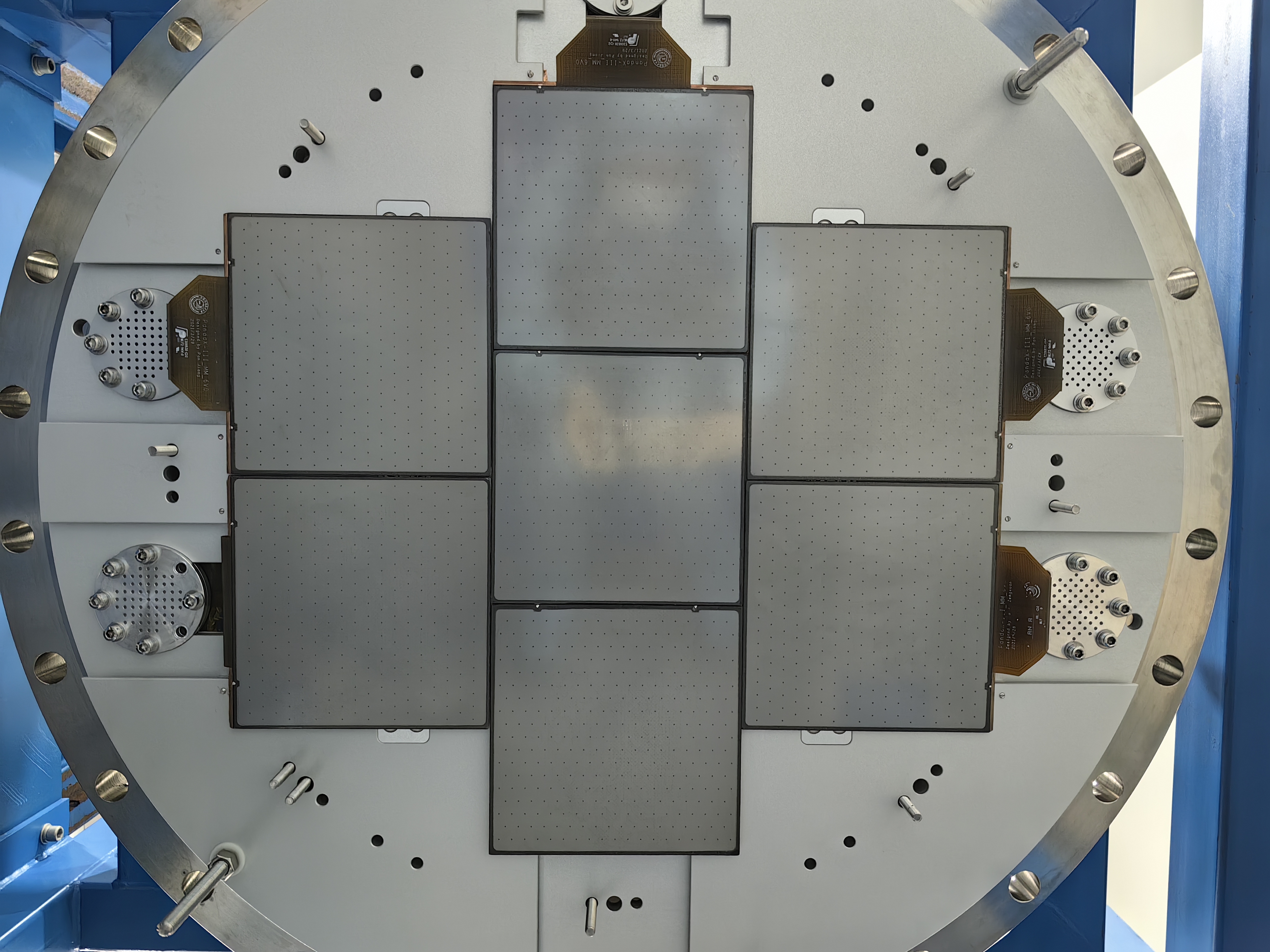}
    }    
    \subfigure[]{
        \label{multiblock}        \includegraphics[width=0.42\textwidth,trim=50 60 20 50,clip]{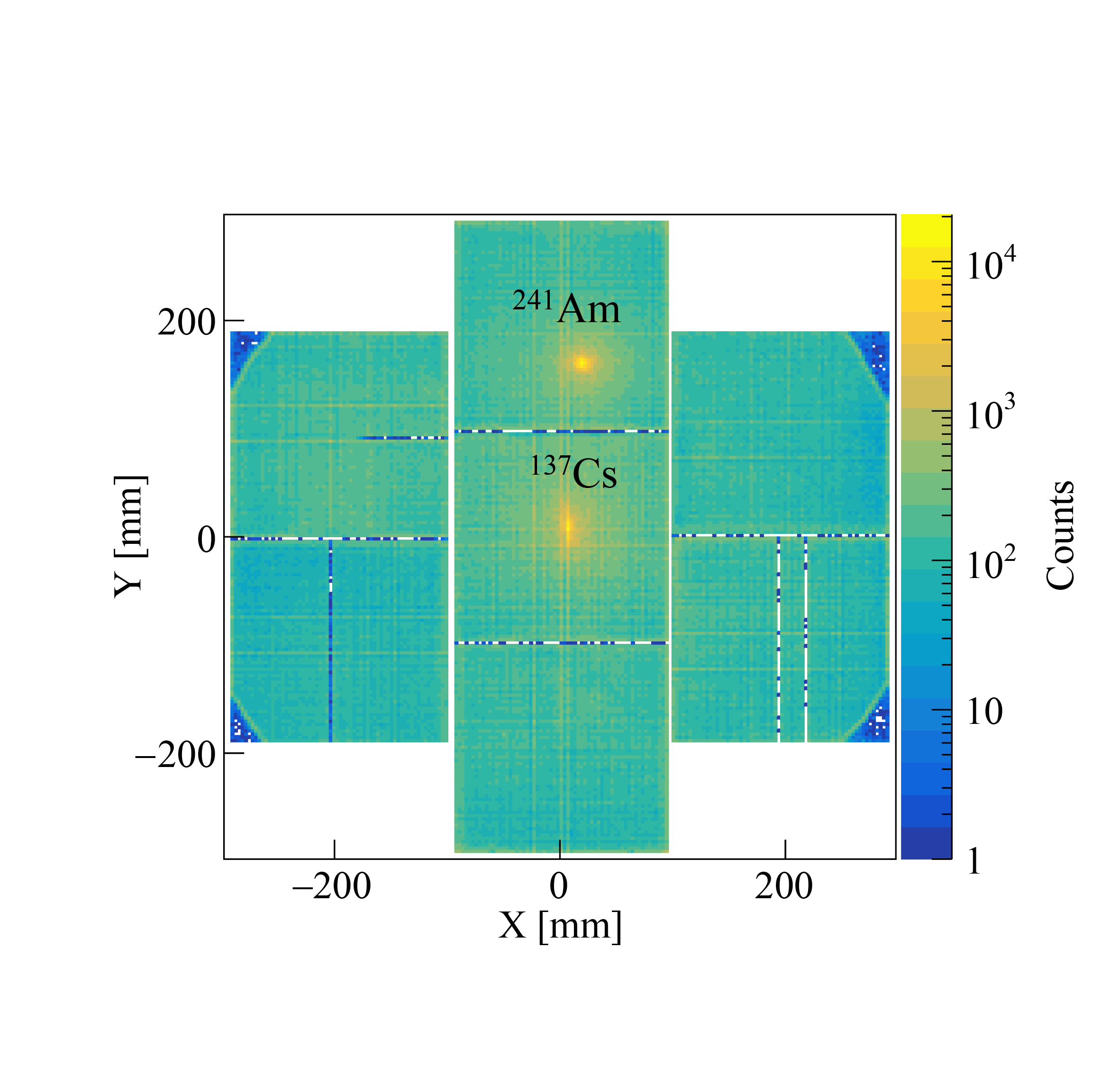}
    }      
    \caption{\label{multiblock_all}
    The readout plane and hitmap for the seven readout modules.
    (a) The readout plane of seven readout modules.
    (b) A hit position map test with seven readout module.
    The bright spot above is the position of the \ce{^{241}Am} source on the XY plane, while the bright spot below indicates the position of the \ce{^{137}Cs} source on the XY plane.}
\end{figure*}

Beyond single-module testing, the DAQ system also supports data acquisition from multiple readout modules. 
To verify its performance under multi-channel conditions, integrated tests were performed using the PandaX-III prototype detector~\cite{b28}, equipped with seven readout modules.
The detector vessel has a volume of 600~L.
The field cage contains a 270~L cylindrical sensitive volume, with a drift length of 78~cm and a diameter of 66~cm. 
The readout plane consists of seven micromegas using screen-printing technology, as illustrated in the Fig.~\ref{readout_plane}.
We placed both \ce{^{241}Am} and \ce{^{137}Cs} sources on the cathode to evaluate the performance of the DAQ software.
The working gas is 10~bar argon mixed with 2\% isobutane.

The Fig.~\ref{multiblock} shows the hitmap results obtained from a \ce{^{137}Cs} source and \ce{^{241}Am} source in seven Micromegas in a long period of time.
The Fig.~\ref{multiblock} illustrates the distribution of trigger positions on the readout plane for electrical signals under a drift field of 1010~V/cm and an amplification field of 97~kV/cm. 
The bright spot above indicates the XY position of source \ce{^{241}Am}, while the bright spot below denotes the XY position of source \ce{^{137}Cs}. 
The lower event rates at the four corners arise from the distorted electric field near the edge, which is caused by the finite size of the field cage.
The white strips in the centre of the readout modules represent defective channels on the electronics and adapter board.

The experimental results show that the DAQ software based on the MIDAS framework works stably in conjunction with the detector, enabling accurate acquisition and processing of experimental data to meet experimental requirements.

\section{Conclusion}\label{sec.VI}

In this work, we have developed a MIDAS-based DAQ software specifically designed for gaseous detectors employed in particle and nuclear physics experiments. This software provides an end-to-end solution covering the entire data lifecycle, including data collection, decoding, storage, and analysis. Equipped with a user-friendly visual interface, it facilitates experimental operation and real-time monitoring of system status. To enhance operational flexibility, the software supports multiple parameter-configuration methods, enabling researchers to customize experimental setups according to their specific requirements.
Upon integration with the REST framework, the software acquires robust data processing and analysis capabilities, allowing decoded and processed data to be stored as variables in ROOT files. Its flexibility overcomes the limitations of traditional DAQ solutions, providing a unified and stable data acquisition platform for gaseous detectors while effectively eliminating compatibility-related issues.

Extensive performance evaluations using two different electronics systems of the PandaX-III experiment have validated the reliability of the developed DAQ system. Using a signal generator, the maximum event rate for a single AGET full readout was measured at 230~Hz, with a corresponding data transfer rate of 15.2~MB/s. The DAQ software has maintained stable operation for 30 days with negligible data packet loss or corruption.

The practical utility of this DAQ system is further demonstrated by its successful application in detector operations. Comprehensive tests were conducted under various experimental conditions using \ce{^{37}Ar} source, including different channel compression settings, sampling rates, and trigger delay times. Electronic noise and self-calibration signals were also evaluated in detail. Furthermore, the data acquisition system was utilized to measure the hitmap and energy spectrum of both the \ce{^{37}Ar} source and the \ce{^{109}Cd} source with a single readout module. The system has also been validated to work reliably with seven readout modules, confirming its effectiveness and suitability in experiments.

\section*{Acknowledgment}

We gratefully acknowledge Changqing Feng, Zengxuan Huang, and Jianguo Liu for their valuable assistance in the joint debugging of the DAQ system and electronic components, and Zhiyong Zhang and Yunzhi Peng for providing the Micromegas detectors for the joint tests.

\end{document}